\documentclass[iop,numberedappendix]{emulateapj}
\newcommand{\be}{\begin{equation}}
\newcommand{\ee}{\end{equation}}

\usepackage{multirow}
\usepackage{hyperref}
\usepackage{natbib, xfrac}
\usepackage{enumerate}
\usepackage{graphicx}
\usepackage{amsmath, amssymb}
\usepackage{courier}
\usepackage{color}

\begin{document}
\title{The Coupled Physical Structure of Gas and Dust in the IM Lup Protoplanetary Disk}
\shorttitle{}
    \shortauthors{Cleeves et al.}

 \author{L. Ilsedore Cleeves\altaffilmark{1,2}, Karin I. {\"O}berg\altaffilmark{1}, David J. Wilner\altaffilmark{1}, Jane Huang\altaffilmark{1}, Ryan A. Loomis\altaffilmark{1},  Sean M. Andrews \altaffilmark{1}, and Ian Czekala\altaffilmark{1}}

\altaffiltext{1}{Harvard-Smithsonian Center for Astrophysics, 60 Garden Street, Cambridge, MA 02138}
\altaffiltext{2}{Hubble Fellow, ilse.cleeves@cfa.harvard.edu}

\begin{abstract}
The spatial distribution of gas and solids in protoplanetary disks determines the composition and formation efficiency of planetary systems. A number of disks show starkly different distributions for the gas and small grains compared to millimeter-centimeter sized dust. We present new Atacama Large Millimeter/Submillimeter Array (ALMA) observations of the dust continuum, CO, $^{13}$CO, and C$^{18}$O in the IM Lup protoplanetary disk, one of the first systems where this dust-gas dichotomy was clearly seen. The $^{12}$CO is detected out to a radius of 970 AU, while the millimeter continuum emission is truncated at just 313 AU.  Based upon this data, we have built a comprehensive physical and chemical model for the disk structure, which takes into account the complex, coupled nature of the gas and dust and the interplay between the local and external environment. We constrain the distributions of gas and dust, the gas temperatures, the CO abundances, the CO optical depths, and the incident external radiation field. We find that the reduction/removal of dust from the outer disk exposes this region to higher stellar and external radiation and decreases the rate of freeze-out, allowing CO to remain in the gas out to large radial distances. We estimate a gas-phase CO abundance of 5\% of the ISM value and a low external radiation field ($G_0\lesssim4$). The latter is consistent with that expected from the local stellar population. We additionally find tentative evidence for ring-like continuum substructure, suggestions of isotope-selective photodissociation, and a diffuse gas halo. \end{abstract}


\section{Introduction} 
With a growing number of high resolution observations, it is becoming increasingly clear that the gas and dust components of protoplanetary disks have significantly different spatial distributions. The emission tracing the ``large'' millimeter sized grains are typically far more radially compact than the molecular gas disk as traced by CO or small grains from scattered light \citep[e.g.][]{pietu2007,isella2007,panic2009,andrews2012,rosenfeld2013a,degregorio2013,walsh2014,huang2016}.  Millimeter-sized grains have also been observed to be trace vertically flatter distributions compared to the flared CO disk and small grains \citep{guilloteau2016,pinte2016}, consistent with earlier results demonstrating a dust mass deficit in the the disk surface layers \citep[e.g.,][]{furlan2006}. These observed morphologies are attributed to dust growth and subsequent settling to the midplane \citep[e.g.,][]{goldreich1973,weidenschilling1980,cuzzi1993,dullemond2004} and preferential radial drift of large grains toward the inner disk \citep[e.g.,][]{whipple1972,weidenschilling1977}. However, why both settling and drift ``halt'' is still an open question \citep[e.g.,][]{birnstiel2014,pinte2014,andrews2016} that is perhaps in part alleviated by disk substructure \citep{whipple1972}.

Tracking the evolution of both gas and grains is critical for understanding how, when, and where these disks form planetesimals, in particular by regulating the surface density of solids. Furthermore, vertical settling and radial drift of ice-coated dust grains change the local chemical composition. These processes together deplete volatile abundances in the surface/outer disk and enhance it in the midplane/inner disk, thus changing the composition of the disk reservoir from which young planets accrete (Bergin et al. 2016, submitted).

The differential transport of dust also affects the disk chemistry itself. As the grains settle and drift inward, the deficit of dust in the upper and outer disk layers exposes the gas to higher levels of both stellar and external radiation. This additional radiation can alter the dust thermal structure \citep{cleeves2016a} and gas temperatures in the disk atmosphere \citep{gorti2015}. Enhanced FUV irradiation (due to decreased UV opacity) enhances photon-driven chemistry, dissociates molecules, photodesorbs ices, and creates radicals that can build toward more complex species. The local abundance of small grains plays a key role in the charge balance, and thus the disk ionization fraction \citep[e.g.,][]{bai2009}. The total surface area of dust, which is largely set by dust growth, also regulates the rate of freeze-out from the gas phase onto icy grains. 

To characterize dust stratification and its effects on the gas, we need to observe disks at high spatial resolution and develop physical models for their underlying density and temperature structures. In this paper, we present new observations of IM Lup (Sz 82) along with detailed physical and chemical models for its gas (as traced by CO) and dust. We investigate how the spatially distinct dust and gas components affect the observed tracers and inferred physical properties, including disk temperatures, UV irradiation, and CO abundances.

IM Lup is a young \citep[$0.5-1$~Myr;][]{mawet2012} M0 type star associated with the Lupus 2 cloud, at a distance of $d=161\pm10$ pc \citep{gaia}. The star has a bolometric luminosity of $L_* = 0.9~L_\odot$ \citep{hughes1994} and mass of 1~$M_\odot$ \citep{panic2009}.\footnote{Values updated from original literature based on new distance.} 
\citet{vankempen2007} detected a gas disk associated with IM Lup, along with large-scale  ($\gtrsim30''$) CO emission from the parent cloud. \citet{lommen2007} detected bright millimeter continuum from the dusty disk using ATCA at 3.3~mm, providing evidence for grain growth to at least millimeter-centimeter sizes. Based upon scattered light observations with {\it HST} and resolved millimeter continuum observations with the SMA, \citep{pinte2008} modeled the dust disk and found that the gas+dust disk is massive, $\sim0.1$~$M_\odot$. Such a large mass is surprising given that IM Lup does not appear to be actively accreting \citep{padgett2006,gunther2010}, though there are suggestions that the accretion rate is variable in time \citep{batalha1993,batalha1998,covino1992,salyk2013}. \citet{panic2009}  observed IM Lup in  $^{12}$CO and $^{13}$CO with the SMA, and found that the gas is more extended than the dust, 900 AU versus 400 AU, respectively. Furthermore, they found that the gas has a substantial break in surface density at 400 AU, the edge of the dust disk, motivating this study of the gas and dust components as individual but intrinsically related components.

We have developed a detailed physical and chemical structure of the gas and dust of IM Lup's protoplanetary disk based on new ALMA observations of CO and its isotopologues and 875~$\mu$m dust, complemented by existing observations presented in \citet{oberg2015im}. This paper is laid out as follows: the new ALMA observations are described in Section~\ref{sec:obs} and the modeling framework used to interpret these observations is outlined in Section~\ref{sec:methods}. We present our results in Section~\ref{sec:results} and discuss their interpretation along with some additional findings in Section~\ref{sec:discussion}.

\section{Observations}\label{sec:obs}
\subsection{Data Reduction}

\begin{deluxetable*}{ccccc}[b!]
\tablecolumns{5}
\tablewidth{0pt}
\tablecaption{Line Observations \label{tab:obs}}
\tabletypesize{\footnotesize}
\tablehead{{Transition} & Rest Freq. & Beam (Position Angle) & Disk-Integrated Flux$^\dagger$ & Moment 0 Image RMS ($1\sigma$) \\
                             & (GHz) &  & (Jy km s$^{-1}$) & (mJy~km~s$^{-1}$)}
\startdata
$^{13}$CO $J=3-2$ & 330.588 & $0\farcs42\times0\farcs32$ $(76.6^\circ)$  & $11.5  \pm 1.7$ & 13.5 \\
C$^{18}$O $J=3-2$& 329.331 & $0\farcs40\times0\farcs35$ $(46.7^\circ)$  &$2.7\pm 0.4$  & 13.2  \\
$^{12}$CO $J=2-1$& 230.538 &  $0\farcs55\times0\farcs41$ $(112.3^\circ)$ & $26.0\pm3.9$ & 5.3  \\
$^{13}$CO $J=2-1$& 220.399 & $0\farcs58\times0\farcs43$ $(111.2^\circ)$   &$ 8.2   \pm1.2$ & 6.3 \\
C$^{18}$O $J=2-1$&219.560 &  $0\farcs58\times0\farcs43$ $(112.0^\circ)$  & $1.4 \pm0.2$ & 4.1 \\
\hline
Continuum & 875 $\mu$m &   $0\farcs37\times0\farcs29$ $(47.0^\circ)$  & $0.59\pm0.09$~Jy & 0.21 mJy~beam$^{-1}$ \\
Continuum & 1.3 mm&   $0\farcs54\times 0\farcs40$ $(-68.3^\circ)$  & $0.20\pm0.03$~Jy  &0.14 mJy~beam$^{-1}$
\enddata
\tablecomments{$\dagger$ Within a $7''$ wide box for $^{12}$CO and $^{13}$CO and a $3''$ wide box for C$^{18}$O. Uncertainties calculated from the RMS scatter in the line free channels combined with 15\% calibration uncertainty.} 
\end{deluxetable*}

We present observations of IM Lup carried out with the Atacama Large Millimeter/Submillimeter Array (ALMA) at Bands 6 and 7 targeting lines of CO and its isotopologues. 
The Band 6 CO data was obtained 2014 July 7 (project code ADS/JAO.ALMA\#2013.1.00226; PI: {\"O}berg) with 21 minutes of on source integration. Thirty-one 12-m antennae were used, with baselines spanning 20 to 650 meters. The correlator was set up with thirteen spectral windows (SPWs). Twelve of them, including those targeting $^{12}$CO $J=2-1$, $^{13}$CO $J=2-1$, and C$^{18}$O $J=2-1$, had 0.061 MHz channels and bandwidths of 59 MHz. 
\begin{figure*}[t]
\begin{centering}
\includegraphics[width=1.0\textwidth]{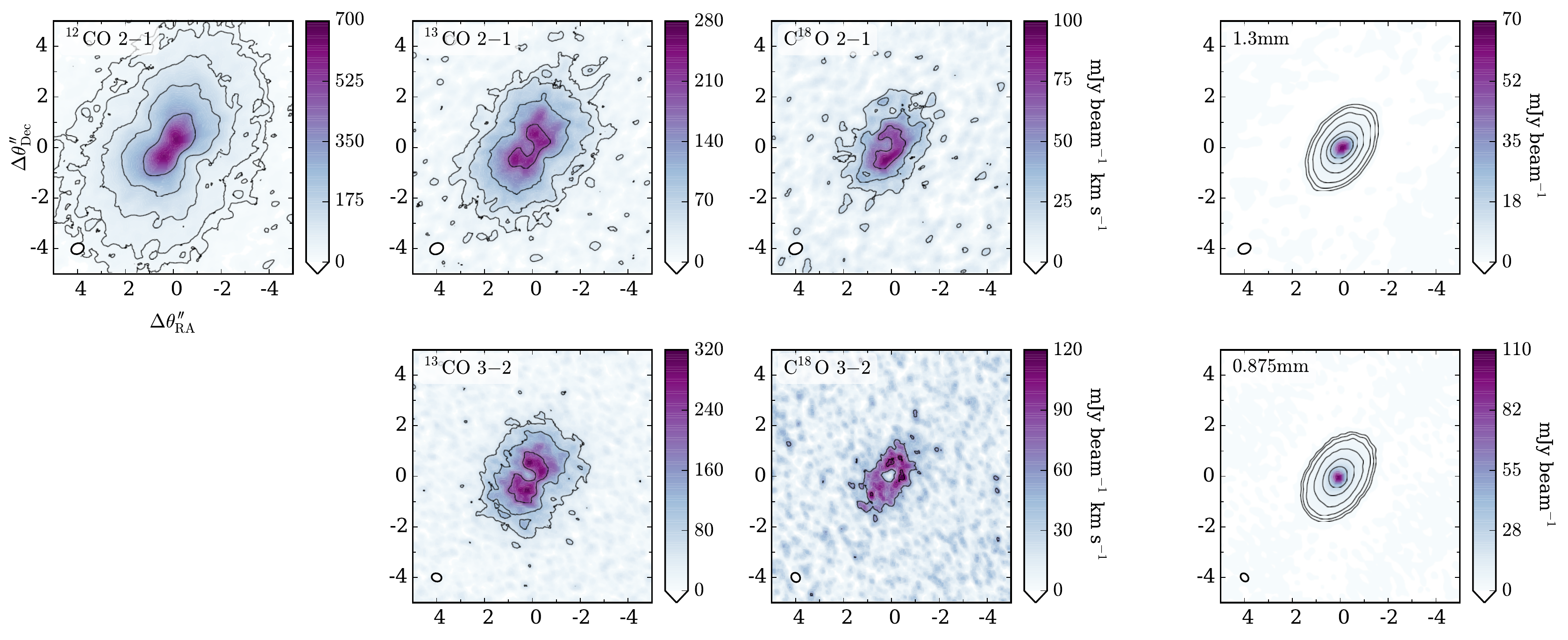}
\caption{Left: Continuum subtracted integrated line intensity for the ALMA Band~6 CO lines (top) and Band~7 lines (bottom). Right: Continuum flux at Band~6 (1.3~mm, top), and Band~7 (875~$\mu$m, bottom). Contours are 4, 8, 16, 48, and 144$\sigma$, where 1$\sigma$ is reported in Table~\ref{tab:obs}. The inner flux deficit visible in the CO isotopologue emission is spatially coincident with the brightest dust emission.  \label{fig:mom0}}
\end{centering}
\end{figure*}
The thirteenth had a channel size of 0.122 MHz and a bandwidth of 469 MHz. 

\begin{figure}[ht]
\begin{centering}
\includegraphics[width=0.4\textwidth]{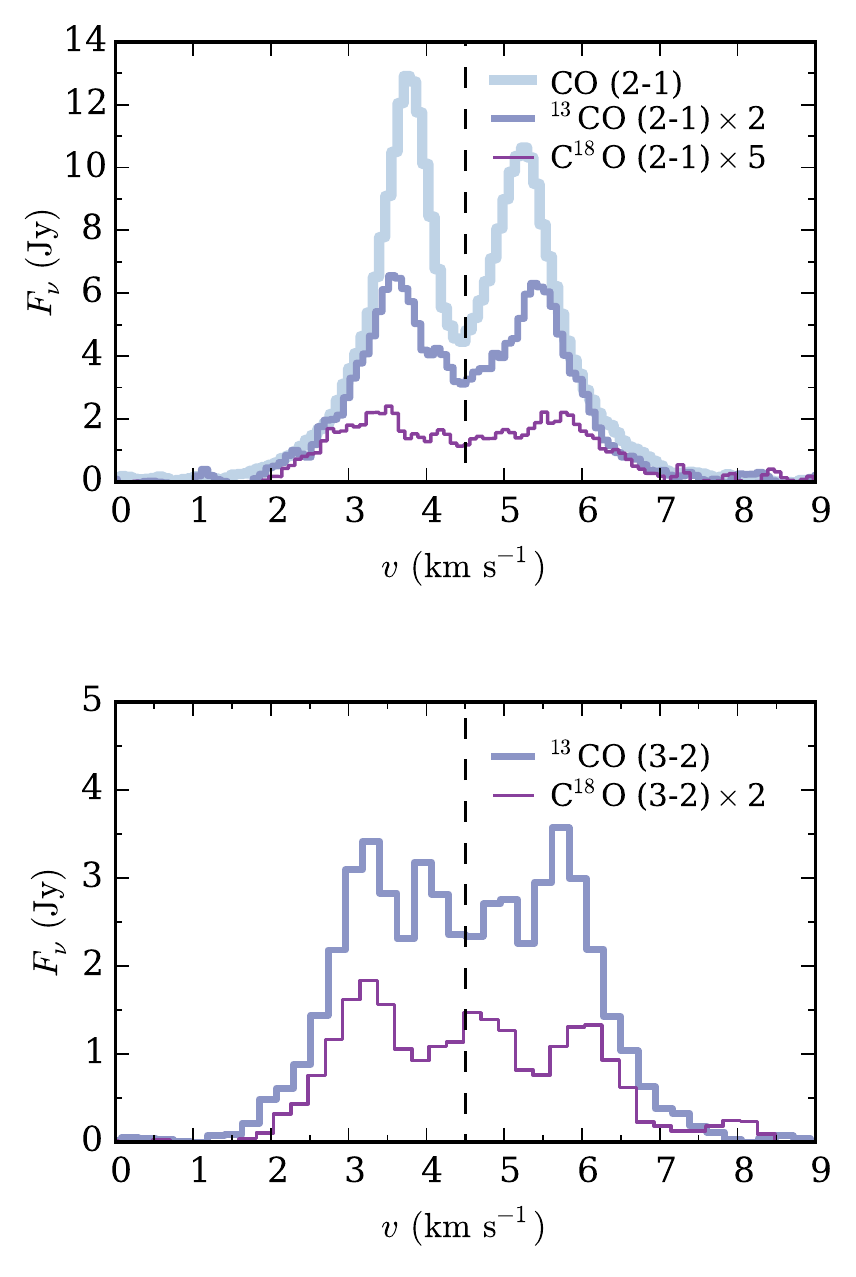} 
\caption{Spatially integrated spectra for the $J=2-1$ transitions (top) and $J=3-2$ transitions (bottom), with scaling factors as indicated. The emission is integrated over an $7''$ box for the $^{12}$CO and $^{13}$CO and $3''$ for C$^{18}$O. The spectra are shown at the resolution of the original observations. \label{fig:spectra}} 
\end{centering}
\end{figure}
The 2013.1.00226 observations were calibrated by ALMA/NAASC staff using the quasars J1427-4206 for the bandpass and J1534-3526 for the phase and amplitude, and Titan for the flux calibration. In addition, we performed one round of phase self-calibration with solutions calculated for each integration (6s) in CASA \citep{mcmullin2007} version 4.4.0. Line-free channels were used to estimate the continuum, which were then subtracted from the SPWs in the uv-plane.   An earlier but similar reduction of the C$^{18}$O $J = 2-1$ data was previously published in \citet{oberg2015im}.

The Band 7 data were observed 2014 June 8 (project code ADS/JAO.ALMA\#2013.00694; PI: Cleeves) with 24 minutes on source and $28 - 646$ meter baselines across 34 antennae. The correlator was configured to target the $J = 3-2$ transitions of C$^{18}$O and $^{13}$CO, along with HC$^{18}$O$^+$ $J=4-3$. The fourth spectral window was set for continuum (channel width of 15.625 MHz). The line-targeted spectral windows had channel widths of 0.244 MHz. We calibrated the observations using the scripts provided by the ALMA/NAASC staff, where the quasar J1517-2422 was used to calibrate the bandpass and J1534-3526 for the phase and amplitude. Titan was used for the flux calibration. One round of phase self-calibration was applied with solution intervals of 30.25 seconds using CASA 4.5.0. The full set of Band 6 and 7 observations are summarized in Table~\ref{tab:obs}. 


\subsection{Observed Features}\label{sec:features}
Figure~\ref{fig:mom0} presents moment 0 maps made using Briggs weighting (robust parameter of 0.5) with multi-scale clean (with scales of 0\farcs03, 0\farcs3, and 0\farcs9). The continuum, also pictured in Figure~\ref{fig:mom0}, is imaged with the same parameters. All of the targeted lines of CO and the continuum were clearly detected and resolved with high signal to noise. The HC$^{18}$O$^+$ $J=4-3$ was not detected at the 4~mJy~beam$^{-1}$ RMS noise level in averaged 0.5 km~s$^{-1}$ channels, and will be discussed in a subsequent paper. The disk integrated spectra for the CO detections are provided in Figure~\ref{fig:spectra}, measured within a $7''$ wide box for the lines of $^{12}$CO and $^{13}$CO and a $3''$ wide box for the C$^{18}$O lines. The observed $^{12}$CO asymmetry was also noted in \citet{panic2009} and attributed to foreground cloud absorption. Figure~\ref{fig:allchannel} shows channel maps in $0.5$~km~s$^{-1}$ wide channels. The signature butterfly pattern of a Keplerian disk is clearly visible  for all of the CO lines. The isotopologue emission appears symmetric in the blue- and red-shifted components. $^{12}$CO $J=2-1$, however, shows a deviation from blue/red-shifted symmetry at low velocities, $\pm0.5$~km~s$^{-1}$ from the systemic velocity of 4.5~km~s$^{-1}$, consistent with the asymmetric spectrum. 

The continuum subtracted CO moment 0 maps in Figure~\ref{fig:mom0} all present flux deficits at the continuum peak location. $^{12}$CO is the least affected, while the C$^{18}$O $J=3-2$ emission drops nearly to zero. To exclude the possibility that these holes are continuum subtraction artifacts, we examined the individual $^{13}$CO $J=3-2$ channel maps without continuum subtraction. We find that the CO emission per channel is not connected at the center, suggesting that the observed emission holes are likely real. We have also estimated the opacity of the $^{13}$CO $J=3-2$ and C$^{18}$O $J=3-2$ emission, since subtraction artifacts are mainly expected for optically thick lines. We created a map of the ratio of the spectrally integrated $^{13}$CO  and C$^{18}$O lines (using the same $0\farcs4$ restoring beam) and find an upper limit to the $^{13}$CO $J=3-2$ optical depth at the continuum peak location of $\tau\lesssim2$. The C$^{18}$O emission is correspondingly optically thin ($\tau\lesssim0.3$). This low line optical depth is inconsistent with a scenario where the CO emission hole is purely a continuum subtraction effect. Instead we consider whether the hole can be attributed to either 1) a true deficit in CO, or 2) continuum opacity ``blocking'' a substantial amount of CO emission {\em above} the disk midplane. 

Scenario 1) appears unlikely when considering that the large amount of dust close the star \citep{pinte2008}, which should protect CO from dissociation. Scenario 2) would require high inner disk dust optical depths at submillimeter wavelengths, which have been inferred for other disks \citep[e.g.,][]{hogerheijde2016}. Because settling time-scales are short, the millimeter dust is often thought to exist in a thin layer near the midplane. However, a thin layer would only block half of the observable CO from the opposite side of the disk and the observed C$^{18}$O emission depth is greater than this. Consequently, scenario 2) requires vertically elevated millimeter grains, or rather high altitude millimeter-wavelength continuum opacity. Below we demonstrate that in light of our new data, this scenario 2) is plausible for IM Lup.

The {\texttt{clean}ed Band 6 and Band 7 IM Lup continuum images show a bright central continuum peak, and fainter emission which extends out as a broad plateau. The outer disk appears truncated at $\sim2''$, i.e. the flux drops by $\sim16\sigma$ over a beam width size scale. We see hints of a faint ring-like structure (weak brightness variations at the $\sim2-3$\% level) in the higher resolution Band 7 continuum, but these features are not directly visible in Figure~\ref{fig:mom0} (see Section~\ref{sec:rings}). 

Using the Band 7 continuum image, we measure a disk position angle of $144\pm3^\circ$ and an inclination of $i=48\pm3^\circ$ (deconvolved from the beam) using the Gaussian fitting tool in CASA viewer. Based upon the brightness and projection of the CO channel maps \citep{rosenfeld2013b}, the southwest quadrant of the image corresponds to the near side of the disk to the observer. Our inclination is in agreement with the value derived by \citet{pinte2008}, $50^\circ$, but in 2$\sigma$ conflict with the value derived by \citet{panic2009} from the gas, $54\pm3^\circ$. This difference may be attributed to the flared CO surface (resolved in our channel maps in Figure~\ref{fig:allchannel}), which intrinsically skews the derived inclination toward more edge-on values. Assuming an $i=48\pm3^\circ$ and a distance of 161~pc, we confirm with the position-velocity diagram tool in CASA that the $^{13}$CO $J=3-2$ rotation profile is consistent with a $1.0\pm0.2$~$M_\odot$ star.

\begin{figure*}[ht!]
\begin{centering}
\includegraphics[width=1.02\textwidth]{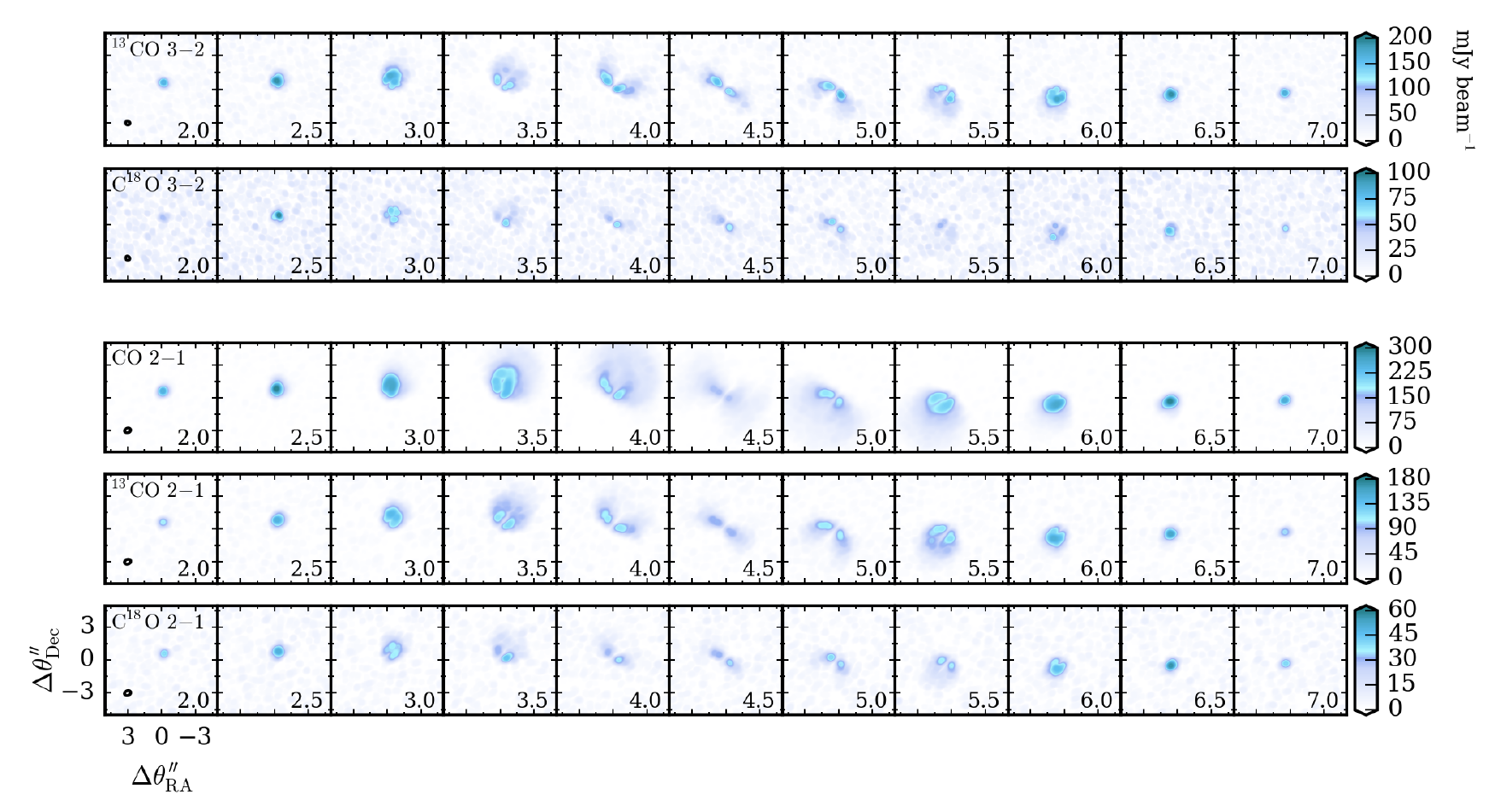}
\caption{Central channel maps for the CO line detections. The beam is in the lower left corner. The VLSR in km~s$^{-1}$ is indicated in the bottom right corner of each panel.  \label{fig:allchannel}}
\end{centering}
\end{figure*}

\section{Methods}\label{sec:methods}

We use the full set of CO and continuum ALMA visibilities, along with the dust spectral energy distribution (SED), to develop our IM Lup disk model. We have recomputed the weights of the ALMA measurement sets using the {\texttt{statwt}} task in CASA, which reweights the visibilities according to the intrinsic scatter in the data. We do not include the low velocity (i.e. central 5 channels) asymmetric $^{12}$CO emission in the analysis, because of potential foreground contamination \citep{panic2009}. Nor do we attempt to model the tentatively detected dust ring structure, but rather assume a smooth dust surface density in the outer disk. 

\begin{figure*}[ht!]
\begin{centering}
\includegraphics[width=1.00\textwidth]{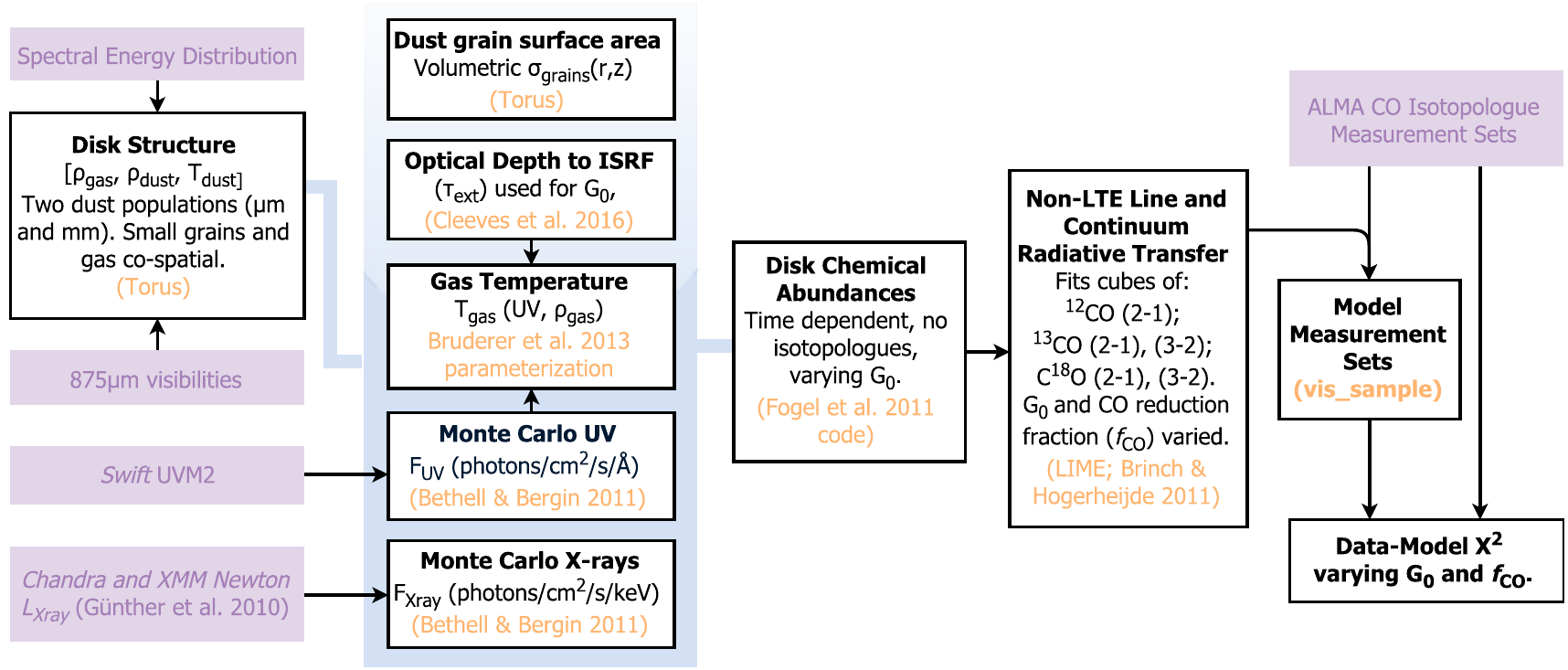}
\caption{Schematic of our disk chemical modeling procedure. Observations are highlighted in purple while methods are shown in orange. \label{fig:method}}
\end{centering}
\end{figure*}

The high quality ALMA data motivates the comprehensive modeling approach outlined below. In particular, we model the dust and gas populations as separate, but connected. We take into account the effects of dust reduction on the gas physics and CO chemistry. The general approach is outlined graphically in Figure~\ref{fig:method} and summarized here. The ALMA Band 7 submillimeter continuum emission provides the structural framework for the large grains, while the distribution of small grains are constrained by the SED and guided by the \citet{pinte2008} scattered light modeling (Section~\ref{sec:physmod}). We assume the gas and small grains are co-spatial, and calculate the X-ray and UV radiative transfer with Monte Carlo radiation transfer (Section~\ref{sec:radmod}). We estimate gas temperatures from the combined stellar and interstellar UV fluxes (Section~\ref{sec:radmod}). These models are passed to the chemical solver to compute the CO abundance versus time (Section~\ref{sec:chemmod}). These abundances along with the global physical structure are used to calculate the emergent line intensities, which are directly compared to observations in the visibility plane (Section~\ref{sec:linemod}). Within this framework, we explore models with 1) varying amounts of gas phase CO depletion, and 2) varying degrees of external FUV radiation contributing primarily to CO dissociation/photodesorption and gas heating. 


\subsection{Disk Structure Model Framework}\label{sec:physmod}

The general form of the physical structure follows \citet{andrews2011}, reflective of the self-similarity solutions of \citet{lyndenbell1974}. The gas surface density is described by:
\begin{equation}\label{eq:sig}
\Sigma_{g} (R) = \Sigma_{c} \left( \frac{R}{R_c}  \right) ^{-\gamma} \exp{ \left[ -\left( \frac{R}{R_c} \right)^{2-\gamma} \right]},
\end{equation}
between an inner and outer radial boundary, $R_{\rm inner}$ and $R_{\rm outer}$, respectively. The critical radius $R_c$ describes the disk location where the surface density transitions from a power law to exponential in radius. $\Sigma_c$ is the characteristic surface density, and $\gamma$ is the gas surface density exponent.  The vertical distribution of gas is described by the gas scale height at 100 AU ($H_{100}$), and the disk flaring parameter ($\psi$). 

We model the dust disk using two grain size populations, a large and small grain distribution. The mass fraction in large grains is parameterized with $f_{\rm mm}$. Both grain populations follow an MRN size distribution \citep{mrn1977} with a minimum grain size of 0.005~$\mu$m and a maximum grain size of 1 $\mu$m and $1$~mm for the small and large grains, respectively. For the dust optical properties we use a mix of 80\% astronomical silicates \citep{draine1984} and 20\% graphite. We assume the gas and small grains are spatially co-located. 

Compared to the gas, large grains are distributed with a smaller vertical scale height and a compacted radial distribution. The vertical distribution is set by the parameter $\chi_{\rm bulk}$, the ratio of the large grain and gas scale heights. The large grain radial distribution is set by a different power law compared to the gas, and an outer ``truncation'' radius, $R_{\rm out, mm}$. $R_{\rm out, mm}$ is smaller than $R_c$, and the large grains thus essentially follow a truncated power law, $\gamma_{\rm mm}$, similar to what was found in TW Hya \citep{andrews2012}. In addition, motivated by the bright inner continuum peak, we have added an inner disk excess in millimeter grains, which is described by a size, surface density enhancement, scale height and power law index. 

The disk-integrated mass in gas is calculated assuming a global gas to dust mass ratio of 100, typical of the ISM \citep{bohlin1978}. Because we adopt a truncated radial distribution of large grains, the local dust to gas ratio varies across the disk. This feature is different from previous work, e.g., \citet{cleeves2015tw}, where we fixed each radius to have $\Sigma_{\rm gas} / \Sigma_{\rm dust} = 100$.

When optimizing the model, we vary all of the small grain/gas and large grain structural parameters except for the inner gas and dust disk edges, and the size of the inner disk. We fix the disk inner edge to $R_{\rm inner}=0.2$~AU \citep{pinte2008}. The size of the inner disk excess component is set to the maximum size that is unresolved by the beam, $R_{\rm inner disk}=21$~AU. 

The disk gas and dust density and dust thermal structure models are computed using the code TORUS \citep{harries2000,harries2004,kurosawa2004,pinte2009}. Based on a given density structure and stellar parameters, we calculate the disk thermal structure in radiative equilibrium using the Lucy method \citep{lucy1999} assuming the disk is passively heated. For the central star, we assume an effective temperature of $T_{\rm eff}=3900$~K and stellar radius of $2.5~R_\odot$ \citep{pinte2008} and a stellar mass of 1~$M_\odot$ \citep{panic2009}, which we hold fixed. We assume a distance of 161~pc based on the recent {\it Gaia} parallax measurement \citep{gaia}, which is closer than what has been adopted in previous modeling of this source \citep{pinte2008,panic2009}.

TORUS outputs synthetic continuum images and spectral energy distributions (SEDs), which we compare to the observed values as described below. For the resolved images, we model the higher resolution Band 7 (875~$\mu$m) continuum visibilities. We compare the model directly in the visibility plane using the \texttt{vis\_sample} package, which is publicly available on github\footnote{\url{https://github.com/AstroChem/vis_sample}}. Using the same spatial frequencies as originally observed, we sample the model images and then compare the goodness of fit with the deprojected, azimuthally averaged visibilities versus spatial frequency.  For the SED fitting, we compare the model to the observed values compiled by \citet{pinte2008}, to which we have added a 6.8~mm observation from \citet{lommen2010} and our 875~$\mu$m and 1.3~mm ALMA data. 

Upon finding a reasonable match to the combined 875~$\mu$m visibilities and SED, we hold the physical structure of the gas and dust disk constant for the remainder of the modeling. Note that the constraint on the dust mass is strong, while for the gas we simply assume that the gas mass is 100 times more massive than the dust mass as we do not have an independent constraint for the gas mass, such as HD \citep{bergin2013hd}.


\subsection{Radiation Field and Gas Temperature Estimates}\label{sec:radmod}

The high energy radiation field drives the disk chemistry, especially at low temperatures, and sets the disk gas temperature in the upper atmosphere where dust and gas temperatures are decoupled. Based upon our best fit model for the dust, we compute the model FUV flux ($912-2000$~\AA) and X-ray flux ($1-20$ keV) from the central star as a function of 2D position in the disk using the Monte Carlo radiation transfer code of \citet{bethell2011u}. For the X-rays, we adopt the quiescent spectral template of \citet{cleeves2013a} and an X-ray luminosity of $4.3\times10^{30}$~erg~s$^{-1}$ \citep{gunther2010}. For the FUV, the \citet{bethell2011u} code computes the UV continuum absorption and scattering off of dust grains using the dust opacity model from the TORUS calculations and the line radiative transfer of Lyman-$\alpha$, which also scatters resonantly off atomic hydrogen atoms, allowing them to travel further into the disk. 

We adopt the shape of the observed FUV spectrum of TW~Hya presented by \citet{herczeg2002,herczeg2004}, which should be a reasonable match given that both of these sources are low accretors at $\le10^{-9}$~${\dot{M}}_{\odot}$~year$^{-1}$ for TW Hya \citep{alencar2002,herczeg2004,ingleby2013} and $\le10^{-11}$~${\dot{M}}_{\odot}$~year$^{-1}$ for IM Lup \citep{gunther2010}. Both sources have been observed by {\it Swift} using the broadband UVM2 filter (PI: Cleeves), which has a central wavelength of 2246~\AA. We use these data to normalize the model input UV spectrum. Using the {\it Swift} UVOT data products and the \texttt{uvotsource} routine in \texttt{HEAsoft}, the UVM2 flux for TW Hya is 13.4 mJy ($\pm0.01$ mJy statistical uncertainty) and for IM Lup is 0.174 mJy ($\pm0.015$ mJy standard deviation over five observations taken from 01/2015 -- 08/2015). Correcting for the different distances and higher extinction toward IM Lup  -- TW Hya is only 55~pc away and has negligible extinction while IM Lup is 161~pc away and has an extinction $A_{\rm V}\sim0.7$ \citep{gunther2010} or $A_{\rm UV}\sim2.1$ from $A_{\rm UV}/A_{\rm V} = 3$ \citep{mathis1990} -- IM Lup is a factor of $\sim1.1$ brighter at 2246~\AA. Such similar FUV fluxes are consistent the sources' currently low accretion rates. To simulate the IM Lup stellar FUV field we uniformly increase the FUV spectral template derived from TW Hya by this factor.

We use the technique from \citet{cleeves2013a} Appendix~A to calculate the radiation from external FUV radiation field throughout the disk. We quantify the external radiation field in terms of the mean interstellar radiation field, $G_0=1$, which corresponds to $10^8$~photons~cm$^{-2}$~s$^{-1}$ between 6 and 13.6~eV \citep{habing}. Following the procedure in \citet{cleeves2016a}, we compute the absorption optical depth to each location throughout the disk uniformly over all $4\pi$ steradian. By taking the optical depth from all directions, the code computes the ``effective'' optical depth to each point. The external radiation field primarily impacts gas photochemistry, ice photodesorption and UV-driven heating. We explore models with $G_0$ values of 1, 2, 4, 8, and 16.  These values are low compared to model predictions of typical massive clusters \citep[e.g.,][]{adams2010}, but appropriate for the Lupus star forming region, which is  populated by relatively low mass stars \citep[A-type and later;][]{hughes1994}.
 
The gas temperature is estimated using the total FUV field and local gas density following the procedure described in \citet{cleeves2015tw}. The temperatures are estimated with a fitting function calibrated to thermochemical models of \citet{bruderer2013}, which parameterizes the gas versus dust temperature decoupling in the disk atmosphere given a UV flux estimate and a local gas density.


\subsection{Chemical Modeling Procedure}\label{sec:chemmod}

We calculate CO abundances using a full 2D time-dependent chemical code, rather than the more common practice of modeling CO parametrically with a uniform abundance bounded by freeze-out and CO photodissociation \citep[e.g.,][]{qi2011,andrews2012}. Typical values used are $T_{\rm dust} < 20$~K and $N_{\rm H_2}\sim10^{-21}$~cm$^{-2}$ for the freeze-out limit and the dissociation limit, respectively. Our departure from this modeling approach is primarily motivated by the presence of extended CO picked up in the sensitive ALMA images. The observed CO is present out to very large radii, where the disk material is expected to be tenuous, suggesting that CO can persist at detectable levels in regions beyond the characteristic dissociation boundary, $N_{\rm H_2}<10^{-21}$~cm$^{-2}$. Likewise, this region is quite cold, well below the freeze out temperature of CO. The use of a chemical model allows us to explore more sophisticated CO chemistry, where -- for example -- we expect a reduction of dust to slow down the rate of CO freeze-out, leading to more gas phase CO than expected.

The disk structural model provides the environmental conditions ($\rho_{\rm gas}$, $\rho_{\rm dust}$, $T_{\rm gas}$, $T_{\rm dust}$, FUV, and X-rays) with which we calculate the chemical abundances as a function of position and time. We use the chemical code of \citet{fogel2011} with updates as described in \citet{cleeves2014par}, including simple grain surface chemistry.  The reaction network includes 6284 reactions and 665 species \citep{cleeves2014par}. The network is built from the OSU gas-phase network presented in \citet{smith2004}. The reactions considered are two-body and include ion-neutral, neutral-neutral, ion dissociative recombination, photon-driven chemistry, freeze-out, thermal and non-thermal desorption including chemical desorption, and self-shielding of CO and H$_2$. The network does not include isotopologues. Based upon previous calculations of \citet{miotello2014}, we expect the lack of isotopologues to most strongly affect C$^{18}$O, while not affecting $^{13}$CO appreciably ($\lesssim20\%$).

CO, the focus of this analysis, is thought to form mainly in the gas phase. It has a relatively simple chemistry, set by freeze-out and thermal/non-thermal desorption along with self-shielded UV photodissociation. A key new feature of these chemical calculations is that we allow the amount of dust surface area per unit volume to vary as a function of position, consistent with the dust disk model.  This feature allows us to treat the loss of dust from the upper layers and the outer disk (and corresponding increase in the inner disk midplane) consistently. Using the dust mass density and average grain size from the small and large dust populations, we calculate the local dust surface area per unit volume as an input to the chemical code. The dust surface area primarily affects chemical processes such as freeze-out and dissociative recombination with charged grains.  

The chemical models are initialized with ``molecular cloud-like'' initial conditions. The baseline abundances are taken from \citet{fogel2011} with updated sulfur abundances as described in \citet{cleeves2014par}. We have further augmented the initial abundances with the observed median ice abundances toward protostars \citep[][Table 3]{oberg2011spitz}. The full set of species initially present and their abundances are provided in Table~\ref{tab:ini}. 

\begin{deluxetable}{llll}[bh!]
\tablecolumns{4}
\tablewidth{0pt}
\tablecaption{Initial chemical abundances relative to total H atoms. \label{tab:ini}}
\tabletypesize{\footnotesize}
\tablehead{{Molecule} & Abundance & {Molecule} & Abundance }
\startdata
H$_2$ & $5.00\times10^{-1}$ & He & $1.40\times10^{-1}$ \\
HCN & $2.00\times10^{-8}$ & CS & $4.00\times10^{-9}$ \\
CO & $4.33\times10^{-5}$ & HCO$^+$ & $9.00\times10^{-9}$ \\
SO & $5.00\times10^{-9}$ & C$^+$ & $1.00\times10^{-9}$ \\
N$_2$ & $1.00\times10^{-6}$ & NH$_3$ & $8.00\times10^{-8}$ \\
H$_3^+$ & $1.00\times10^{-8}$ & C$_2$H & $8.00\times10^{-9}$ \\
H$_2$O(gr) & $2.40\times10^{-4}$ & CH$_4$(gr) & $5.70\times10^{-6}$ \\
CH$_3$OH(gr) & $1.80\times10^{-5}$ & CO$_2$(gr) & $3.30\times10^{-5}$ \\
N & $2.25\times10^{-5}$ & CN & $6.00\times10^{-8}$ \\
Si$^+$ & $1.00\times10^{-9}$ & Mg$^+$ & $1.00\times10^{-9}$ \\
Fe$^+$ & $1.00\times10^{-9}$ & Grains & $6.00\times10^{-12}$ 
\enddata
\end{deluxetable}

\subsection{Line Simulations}\label{sec:linemod}

To compare our CO abundance calculations to the observations, we simulate the emergent line intensity predicted by the chemical models. From the model abundances, we compute line intensities using the non-LTE line radiative transfer code LIME v1.6 \citep{brinch2010}. The collisional rates were provided by the Leiden LAMDA database \citep{schoier2005}, originally computed by \citet{yang2010} and \citet{jankowski2005}. The gas motions include Keplerian rotation, thermal broadening from the gas kinetic temperature, and a fixed turbulent broadening width of 100~m~s$^{-1}$. The latter is based on observationally derived upper values in TW Hya \citep{hughes2011,teague2016}.  The disk distance and inclination/position angle are fixed to that derived from the continuum. As noted in \citet{rosenfeld2013b}, LIME creates images at a single velocity, and does not take into account spectral averaging within a channel. To address this we simulate the lines at $10\times$ higher resolution than the channel size used for the model/data comparisons, 0.5~km~s$^{-1}$, and average the narrow channels back down to 0.5~km~s$^{-1}$ resolution. For the isotopologues, we adopt fixed abundance ratios of $^{16}$O/$^{18}$O of 540 and $^{12}$C/$^{13}$C of 70 \citep{henkel1994,prantzos1996}. 


\section{Results}\label{sec:results}


\subsection{Gas and Dust Physical Structure}\label{dustmod}
\begin{figure}[t]
\begin{centering}
\includegraphics[width=0.42\textwidth]{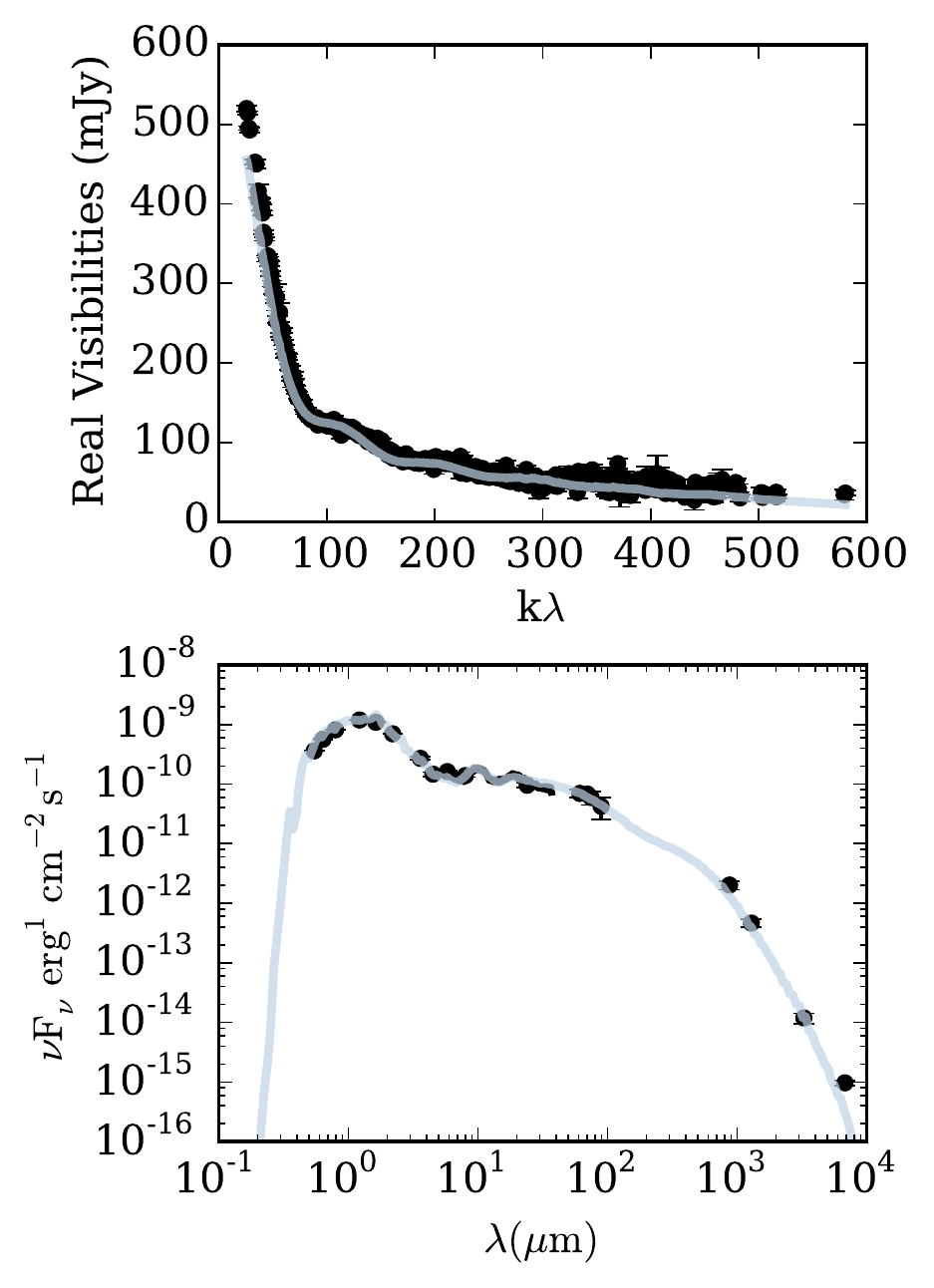} 
\caption{Top: Observed real deprojected visibilities at 875~$\mu$m averaged into 400 evenly spaced bins between 0 and 600 $k\lambda$ with $1\sigma$ errors on the mean (black points). Best fit dust model is shown in gray. Bottom: IM Lup spectral energy distribution tabulated by \citet{pinte2008} from 2MASS, Spitzer, and SMA data, including optical data from \citet{padgett2006} and 3.2~mm from \citet{lommen2007}. Original fluxes are provided in Table~6 of \citet{pinte2008}. The 6.8~mm flux is taken from \citet{lommen2010} and we have added our Band 6 and 7 disk-integrated continuum measurements. The best fit model is overlaid in gray. \label{fig:dustfit}}
\end{centering}
\end{figure}

Using the modeling framework described in Section~\ref{sec:physmod}, we have compared the model SED and 875~$\mu$m visibilities to the observations by eye. Our resulting favored model is compared with the data in Figure~\ref{fig:dustfit} and is described by the parameters listed in Table~\ref{tab:mod}. The total dust mass is $1.7\times10^{-3}$~$M_\odot$, which for a gas to dust mass ratio of 100 corresponds to a gas mass of 0.17~$M_\odot$ within the $R_{\rm out} = 1200$~AU model space. While this is a massive disk, we have confirmed that it is Toomre stable ($Q > 1$) at all radii. The minimum $Q$ value is 3.7 at 70~AU and $\ge6$ when $R\ge200$~AU.

\begin{figure*}[ht!]
\begin{centering}
\includegraphics[width=0.8\textwidth]{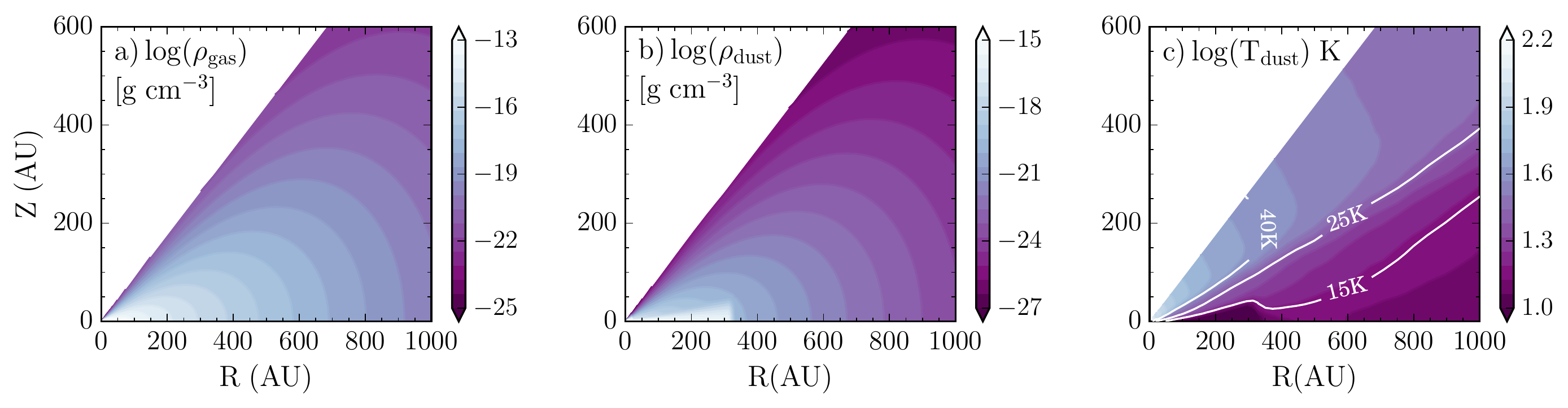} %
\caption{IM Lup model disk structure: a) gas density; b) total dust density from small and large grains; and c) dust temperature. \label{fig:structure}}
\end{centering}
\end{figure*}

We concur with \citet{pinte2008} that disk is quite flared: our estimate of the scale height at 100~AU is 12~AU. A slightly larger scale height ($14$~AU) provides a better match to the 875~$\mu$m flux and visibilities, but a worse match for the SED. The disk is settled, with a best-fit relative scale height for the millimeter grains, $\chi_{\rm bulk}$, of 0.25. This value is slightly less settled compared to other massive disks that are reasonably fit by $\chi=0.2$ \citep{andrews2011}. Furthermore, we find that the large grain disk is outwardly truncated at $313\pm15$~AU based on the slope of the visibilities at short spatial frequencies, which coincides with the location of a reported DCO$^+$ ring \citep{oberg2015im}. 

We did not fit directly to the 1.3~mm visibilities, which enables us to use them as a test of the model structure. Taking our best fit model we have compared the predicted synthetic visibilities to the observations, and find reasonable agreement  with the overall shape and brightness of the visibility profile (within $2\sigma$ for most spatial frequencies). The main difference is that the 1.3~mm visibilities appear to trace a slightly smaller truncation radius than the 875~$\mu$m by a few tens of AU, though still much larger than 250~AU.

\begin{deluxetable}{ccc}[bh!]
\tablecolumns{3}
\tablewidth{0pt}
\tablecaption{Disk Model Fitted Parameters \label{tab:mod}}
\tabletypesize{\footnotesize}
\tablehead{{Parameter} & Value & Definition }
\startdata
$\Sigma_c$ & 25 g~cm$^{-2}$  & Characteristic surface density\\
$R_c$ & 100 AU & Critical radius of gas \\
$H_{100}$ & 12 AU & Scale height for gas/small grains \\
$\psi$ & 0.15 & Flaring parameter \\
$\gamma$ & 1.0 & Gas surface density power \\
$R_{\rm inner}$ & 0.2 AU & Inner disk edge \\
$R_{\rm out, gas}$ & 1200 AU & Gas outer radius \\ 
& & \\
$\chi_{\rm bulk}$ & 0.25 & Relative mm-grain scale height \\
$R_{\rm out, mm}$ & 313 AU & Large grain outer radius \\
$\gamma_{\rm mm}$ & 0.3 & Power law for large grains \\
$f_{\rm mm}$ & 0.99 & Mass fraction in large grains \\
& & \\
$R_{\rm inner disk}$ & 21 AU & Size of inner disk component \\
$F_{\rm inner disk}$ & 4 & Inner surface density enhancement \\
$\chi_{\rm inner}$ & 1 & Inner relative mm-grain scale height \\
$\gamma_{\rm mm, inner}$  & 0.3 & Inner surface density power 
\enddata
\end{deluxetable}

\begin{figure*}[ht!]
\begin{centering}
\includegraphics[width=1.0\textwidth]{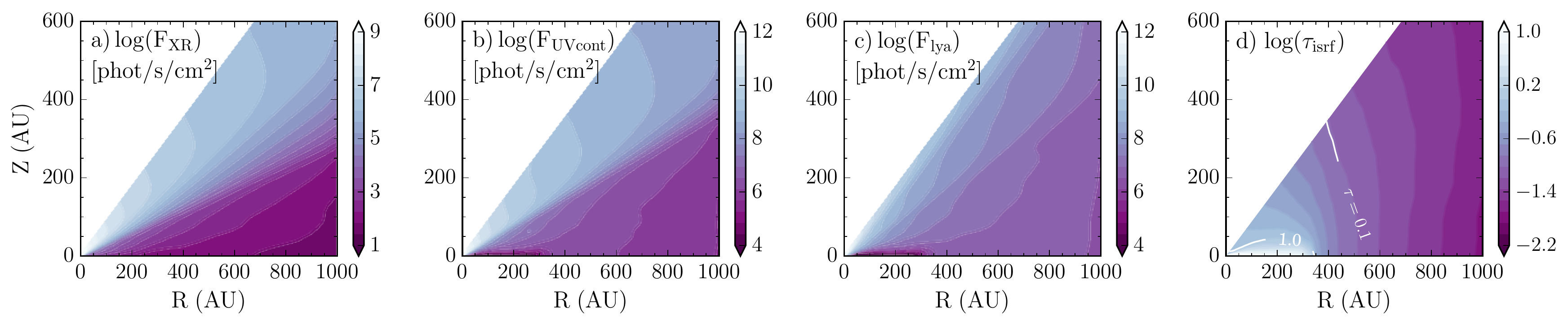} 
\caption{High energy radiation from the central star throughout the disk: a) integrated X-ray flux ($1-20$~keV); b) integrated UV continuum flux ($912-2000 \rm \AA$); c) integrated Lyman-$\alpha$ flux; and d) optical depth to the interstellar FUV radiation field.  \label{fig:internalrad}}
\end{centering}
\end{figure*}

Given the lack of information on small scales at our resolution limit (beam of $\sim50$~AU), the the properties of the bright inner disk are less well constrained.  Exterior to the central dust peak, we can model the broad plateau of 875~$\mu$m flux out to 313~AU with a shallow dust surface density power law of $\gamma_{\rm mm}=0.3$. Extending this to small scales results in an unresolved point-source residual of 26~mJy, or $130\sigma$. This excess central point source is consistent with the observed flattening-out above zero in the deprojected continuum visibilities. To fit this emission, we have included an extra inner disk component as discussed in Section~\ref{sec:physmod}. Based on the CO data, the millimeter grains in the inner disk are likely extended above the geometrical midplane to explain the depth of the CO inner deficit. The emitting region is also compact, $R\le21$~AU, otherwise the emission would start to become resolved in our beam, and the emission is optically thick. Because the inner disk is optically thick and size-limited, the only way to make it brighter is to make it hotter or more massive, or change the opacities. We can fit most of the central brightness by increasing the inner disk mass ($F_{\rm inner disk}$) and scale height of large grains ($\chi_{\rm inner}$) to match the scale height of the gas, which makes the emitting region warmer. The large scale height required to make the inner disk sufficiently warm in turn implies that the inner disk is vertically well-mixed, consistent with the constraints from the CO data. The best-fit model can explain 21 out of 26~mJy central emission excess, leaving 5~mJy residual emission. The resulting inner disk component contains 0.035~$M_\odot$ or about 20\% of the total disk mass. Figure~\ref{fig:structure} presents the underlying density structure for the gas, dust, and dust temperature for our favored model.

\subsection{UV and X-ray Radiation Fields}\label{sec:radresults}

From the disk physical model, we estimate the radiation field throughout the disk (Section~\ref{sec:radmod}).
The model X-ray flux, UV continuum, Lyman-$\alpha$ fluxes, and optical depth to the external radiation field are shown in Figure~\ref{fig:internalrad} as a function of position within the disk. The X-rays are affected by both the gas and dust distribution and smoothly vary with distance from the star and depth into the disk. The UV continuum photons, which are sensitive to the dust distribution, wrap around the large grain disk and are scattered into the outer disk. Similarly, the Lyman-$\alpha$ photons, which are also resonantly scattered by atomic hydrogen in the upper layers and forward scattered by dust below the H -- H$_2$ transition and are able to penetrate deeper than the continuum UV photons in the outer disk. Still the outer disk is dominated by the external radiation field, even for $G_0=1$; Figure~\ref{fig:internalrad}d shows that the optical depth to the external radiation field is only about $\sim1$ at the position of the large grain outer radius.

The stellar FUV field combined with varying levels of $G_0$ produces the different gas temperatures in Figure~\ref{fig:COTG}a. Increasing $G_0$ increases the outer disk midplane gas temperature from a base level of $\sim10-15$~K for a mean ISRF of $G_0=1$ to $\sim15-25$~K for an elevated $G_0=16$. We note that the dust temperature remains fixed to the radiative equilibrium calculated values described in Section~\ref{sec:physmod} regardless of $G_0$.

\begin{figure*}[t]
\begin{centering}
\includegraphics[width=1.0\textwidth]{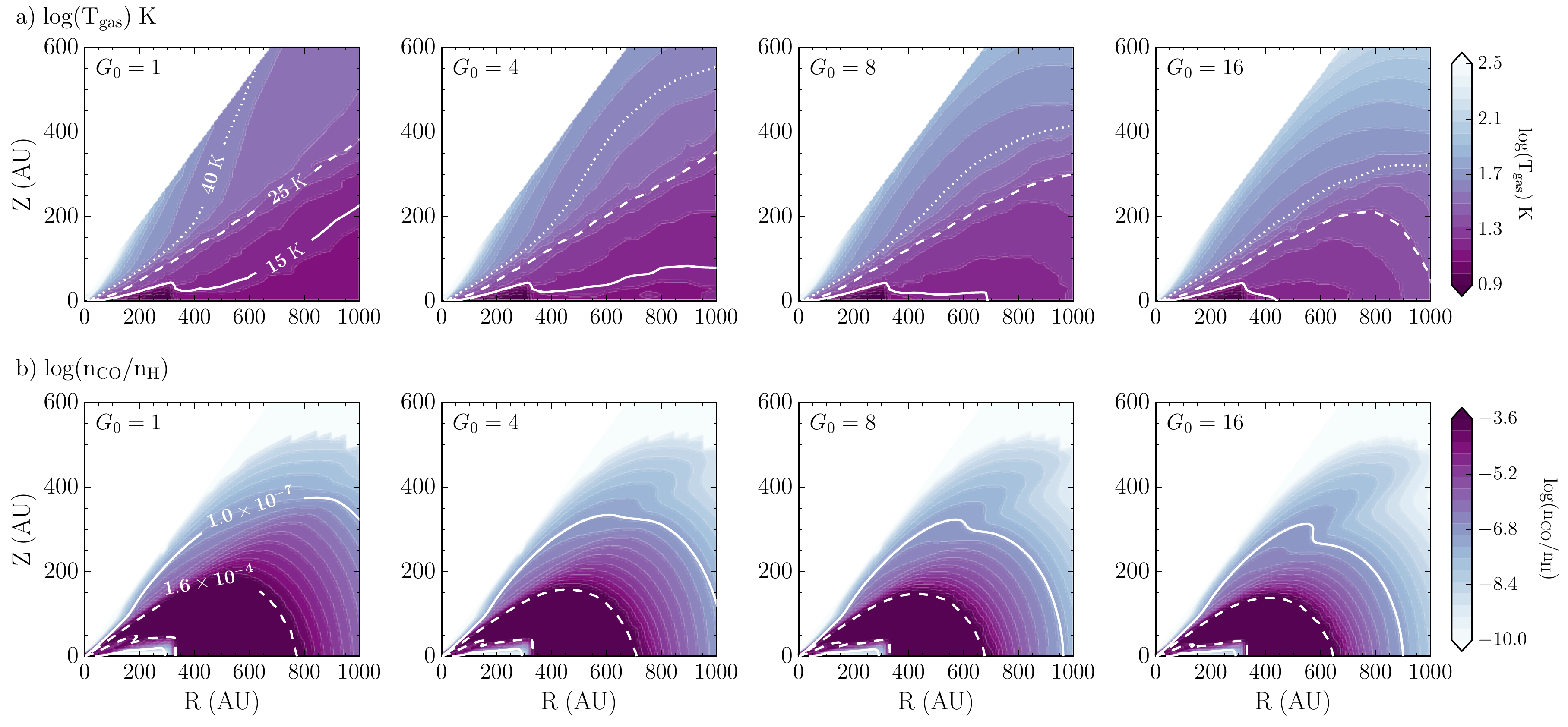} 
\caption{a) Gas temperature calculated from the combined stellar and interstellar UV field, where $G_0=1$ is the average interstellar radiation field. External radiation contributes most significantly to the gas temperatures in the outer ($>500$~AU) disk. The contour lines mark specific  temperatures as labeled in the top left panel. b) Calculated gas-phase CO abundance for models with an increasing external radiation field. The contour lines mark specific CO abundances as labeled in the bottom left panel: High G$_0$s truncate the extent of the CO disk closer inward.  \label{fig:COTG}}
\end{centering}
\end{figure*} 


\subsection{Chemical Model Results}

From the combined physical structure and internal stellar radiation field calculated above, we model the gas-phase CO abundances for a range of external radiation fields. Figure~\ref{fig:COTG}b shows the CO abundance distributions for a subset of external irradiation levels or $G_0$s. Self-shielding preserves gas-phase CO out to radii of many hundreds of AU in all models, but the dissociation region is pushed inward for the higher $G_0$ models by $\sim100$~AU compared to $G_0=1$.

The CO abundance distribution in the warm molecular layer is set by in situ disk chemistry, which quickly drives the initial CO abundance to `typical' values $(1-2)\times10^{-4}$, even though a large fraction of carbon and oxygen start in less volatile CH$_3$OH and CO$_2$ ices (Table~\ref{tab:ini}).  The initial ices are converted to CO via UV desorption, molecular dissociation, and numerous CO reformation pathways, and at 1 Myr,  $\sim25-50\%$ of the available oxygen  is in CO in this layer.

Our models also produce abundant gas-phase CO in the outer disk midplane beyond the large grain radius (313~AU). In the CO column density shown in Figure~\ref{fig:mod_column} (top), there is a $\sim20\%$ jump  at this location. 
Recent modeling in \citet{cleeves2016a} explored the possibility of external CO desorption fronts either due to thermal desorption or photodesorption. While we do see the midplane increase in dust temperature just beyond the large grain radius (see Figure~\ref{fig:structure}), it does not exceed the CO freeze-out temperature. The source of the CO in the outer disk is instead an indirect UV photodesorption pathway.  The cold temperatures combined with relatively high UV causes CO to rapidly form CO$_2$ ice via grain surface reactions. The CO$_2$ ice is then photodesorbed and photodissociated back into CO. This process dominates over direct CO photodesorption. The presence of UV and decreased rate of freeze-out maintains a high level of gas phase CO, $\sim10^{-4}$ relative to total H. Furthermore, combination of photodesorption and CO conversion into CO$_2$ produces overall low ice abundances ($\sim10^{-7}$), and an elevated CO$_2$/CO ice ratio.

\begin{figure}[t!]
\begin{centering}
\includegraphics[width=0.45\textwidth]{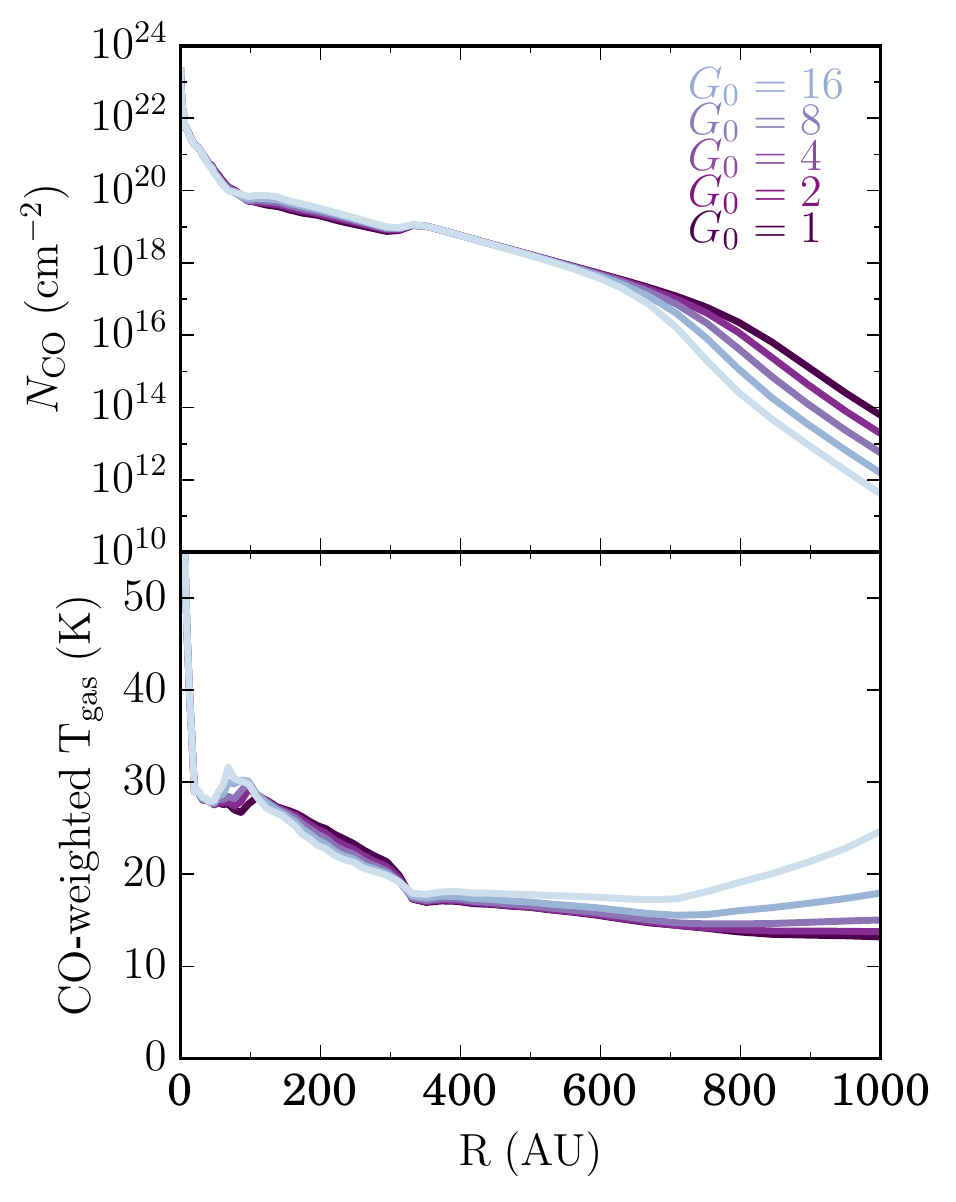} 
\caption{Top panel: CO column density versus radial distance from the star. Bottom panel: CO-weighted gas temperature. High external radiation simultaneously dissociates more CO but leaves the existing CO warmer. \label{fig:mod_column}}
\end{centering}
\end{figure}

As described in Section~\ref{dustmod}, the external FUV also heats the gas. Increasing G$_0$ results in a higher characteristic temperature of CO emitting gas in the outer disk (Figure~\ref{fig:mod_column}). The characteristic or, more precisely, CO-weighted average temperature is calculated by 
\begin{equation}\label{eq:Tg}
T_{\rm avg} ({\rm R}) = \frac{ \int_{0}^{\infty} n_{\rm CO}({\rm R},{\rm Z}) T_g({\rm R},{\rm Z})  d{\rm Z}}{\int_{0}^{\infty} n_{\rm CO} ({\rm R},{\rm Z})  d{\rm Z}},
\end{equation}
where $n_{\rm CO}$ and $T_g$ are the local CO volume density and gas temperature. Beyond the large grain radius, the CO weighted gas temperature decreases with distance from the star for the low ISRF models with $G_0\le4$. The gas temperature reverses and begins to increase with distance for the higher $G_0$ models, increasing by 5~K and 10~K for $G_0$ of 4 and 16 at the disk outer radius, compared to the $G_0=1$ model. At the edge of the large grain disk, 313~AU, the weighted CO gas temperature drops by about $3-4$~K in all models. This drop occurs because at this radius there is excess cold, photodesorbed CO contributing to the CO-weighted temperature. At this same location, the dust temperature increases by about 6~K in the disk midplane, from 5~K to 11~K, due to efficient propagation of reprocessed radiation from the surface layers in this lower density region\citep{cleeves2016a}.


\subsection{Constraints on the CO Abundance and the External Environment}

\begin{figure*}[ht!]
\begin{centering}
\includegraphics[width=1.0\textwidth]{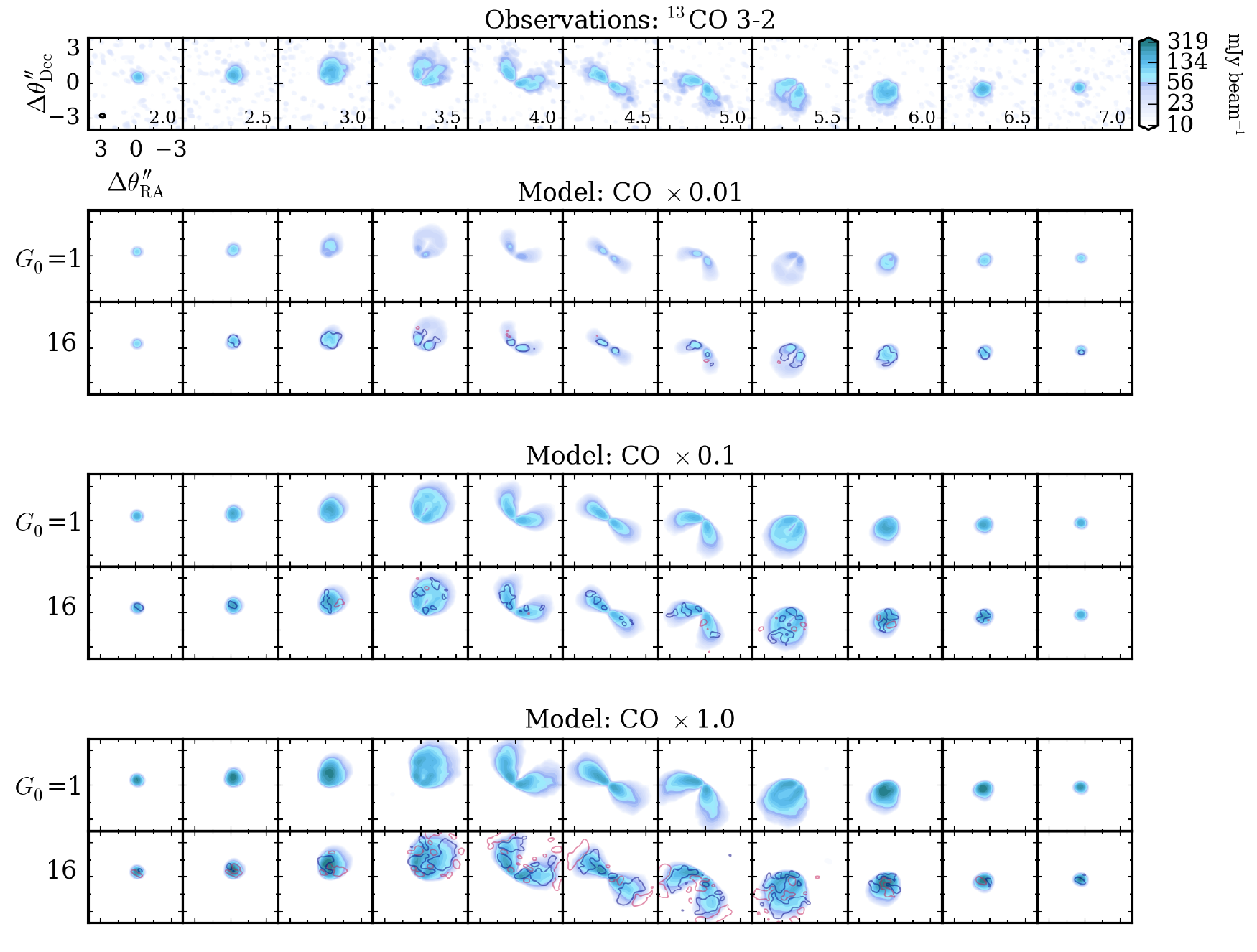}
\caption{Sample continuum subtracted channel maps for $^{13}$CO $J=3-2$ for our grid of models in 0.5~km~s$^{-1}$ channels. Models for three CO reduction factors (0.01, 0.1 and 1.0) are shown from top to bottom. For each set, the highest and lowest $G_0$ considered are shown. The CO reduction factor is the dominant factor, while $G_0$ affects the $^{13}$CO emission more subtly. As is most clearly seen in the models with no CO reduction (bottom), the high $G_0$ field compared to the low truncates the outer edge of the CO and makes the CO emission brighter at the limbs of the wings. Contour lines on the bottom panel show the $1\sigma$ residual emission between the $G_0=1$ and $G_0=16$ cases where blue is positive and red is negative. \label{fig:imlupsample}}
\end{centering}
\end{figure*}

To compare our CO chemical models to the ALMA observations, we follow the procedure outlined in Section~\ref{sec:linemod}. In addition to varying $G_0$, we also explore the effects of a uniform CO reduction factor ($f_{\rm CO}$), applied directly to the calculated CO abundances prior to the LIME line radiative transfer simulations. We examine models with six different values of $f_{\rm CO}$, 0.01, 0.05, 0.1, 0.2, 0.5, and 1.0 (no reduction). This second variable is motivated by evidence of low CO abundances in other disks, $1-2$ orders of magnitude below the canonical value of $\sim10^{-4}$ \citep{favre2013,kama2016,mcclure2016}. Formally, variations in $f_{\rm CO}$ either trace a reduction in gas mass or a reduction in CO abundance, and without an additional gas tracer such as HD these scenarios are difficult to disentangle \citep{bergin2013,mcclure2016}.

Figure~\ref{fig:imlupsample} presents an illustrative subset of the line model grid for the $^{13}$CO $J=3-2$ transition. The dominant factor in our models is the CO reduction factor. The best fit model has a low CO abundance, but not as low as in TW Hya \citep{favre2013}. $G_0$ has a smaller effect. The low sensitivity to $G_0$ is a result of the interplay of temperature and CO column density from Figure~\ref{fig:mod_column}. Essentially, low $G_0$ models produce more CO at large radii, but the CO is on-average significantly cooler and therefore emits less. At the other extreme, a high $G_0$ produces very warm CO, but $1-2$ orders of magnitude lower column density beyond 700~AU because of enhanced photodissociation. 

\begin{figure*} 
\begin{centering}
\includegraphics[width=1.0\textwidth]{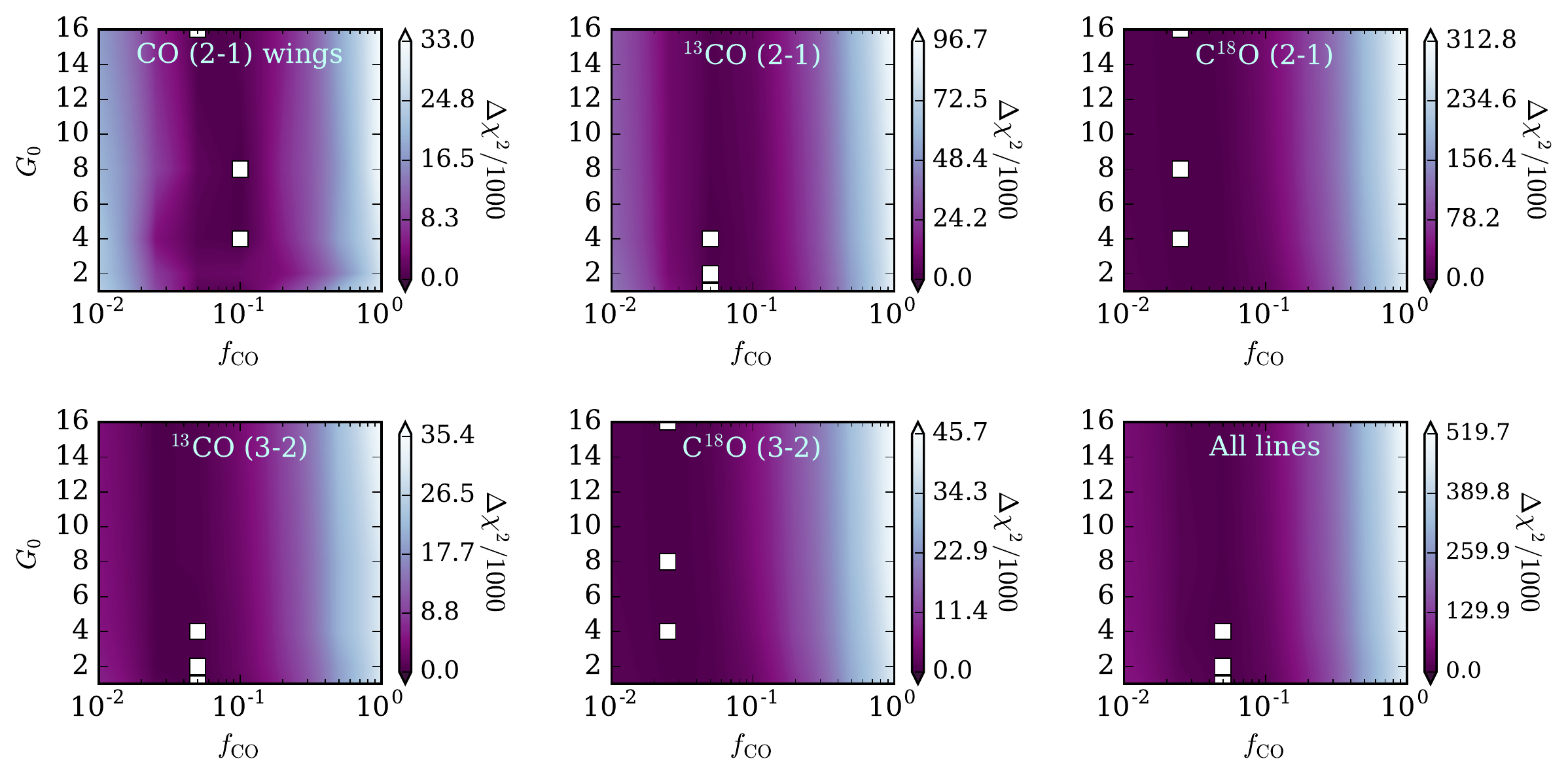}
\caption{$\Delta\chi^2=\chi^2 - {\rm min}(\chi^2)$ for the grid of models varying the CO reduction factor ($f_{\rm{CO}}$) and $G_0$ for all of the detected CO lines. Data and models are compared in the visibility plane after continuum subtraction. The bottom right panel shows the global  $\Delta\chi^2$ for all lines. The white squares highlights the three best fit regions (of the total 25 models) for clarity. \label{fig:chisq}}
\end{centering}
\end{figure*}

\begin{figure*} 
\begin{centering}
\includegraphics[width=1.0\textwidth]{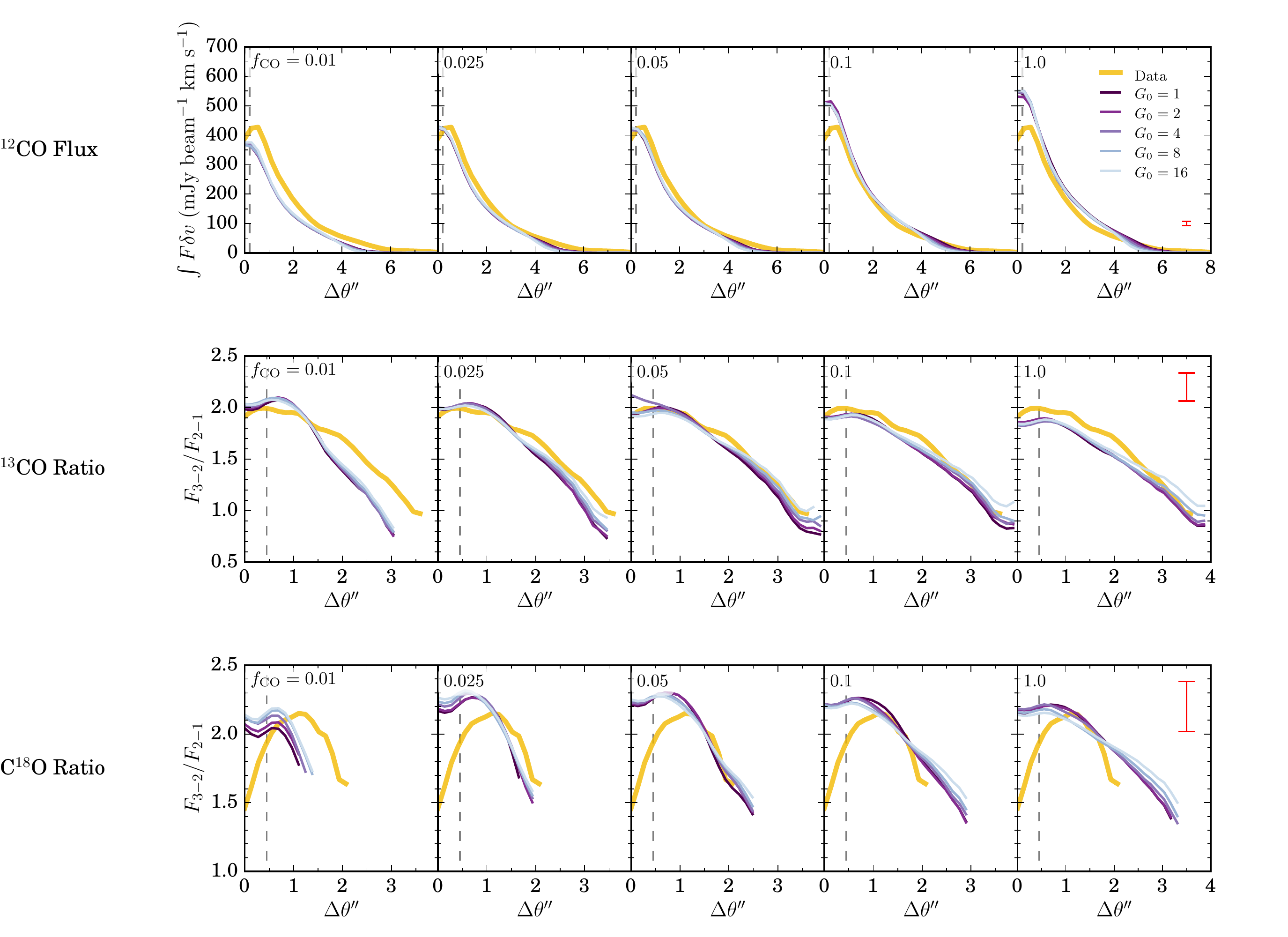}
\caption{Observed, deprojected radial emission profiles and line ratio profiles compared to the models, which provide additional temperature constraints. Top: $^{12}$CO $J=2-1$ integrated line intensity versus radius; middle: the $^{13}$CO $J=3-2$ / $J=2-3$ line ratio versus radius; bottom: C$^{18}$O $J=3-2$ / $J=2-3$ line ratio versus radius. Data are shown in the thick yellow line and models in shades of purple. The $1\sigma$ error on observed line flux or ratio is shown as the red vertical bar on the rightmost panels. The dashed vertical line indicates where the dust is becoming optically thick for Band 6 (top) and Band 7 (middle, bottom) according to the models. \vspace{0.5cm} \label{fig:lineprof}}
\end{centering}
\end{figure*}

Figures~\ref{fig:chisq} and \ref{fig:lineprof} present the model results for the full grid of observed CO lines varying $G_0$ and $f_{\rm CO}$. The $\Delta\chi^2$ in Figure~\ref{fig:chisq} takes into account both real and imaginary visibilities over all channels where the line was detected at greater than $3\sigma$ in the integrated spectrum (Figure~\ref{fig:spectra}). For $^{12}$CO $J=2-1$ we exclude the central 5 channels ($\pm1$~km~s$^{-1}$) in the $\chi^2$ due to possible envelope and/or cloud contamination \citep[see also Section~\ref{sec:diffuse} and ][]{panic2009}. 

The almost vertical regions of lowest $\Delta\chi^2$ confirm that the emission from individual CO lines is more sensitive  to $f_{\rm CO}$ than $G_0$. Nonetheless, most of the observed lines tend to favor low $G_0$, i.e. the best fit solutions have $G_0 \le 4$. The same best fit solution, CO reduction of 20 and $G_0\le 4$, is recovered if models and data are compared without continuum subtraction, indicating that the combined gas and dust model is reasonable. 

To further constrain $G_0$, we exploit the fact that $G_0$ affects the gas temperature (i.e., Figure~\ref{fig:mod_column}). Our multi-line dataset allows us to compare the temperature-sensitive line ratios of both $^{13}$CO and C$^{18}$O, along with the optically thick (and thus temperature sensitive) $^{12}$CO $J=2-1$ intensity profile to the models at different $G_0$.  We emphasize that while the line ratios are sensitive to the temperature profile they are not direct tracers of it; $J=3-2$ and $J=2-1$ lines emit from different disk layers, which may be characterized by different temperatures since the vertical temperature gradient in the disk is steep.

To compute the line ratios we have re-imaged the lines with a uniform restoring beam of $0\farcs6$. We then take the line ratio in the image plane, deproject the disk inclination, and azimuthally average to get the line ratio versus distance, shown in Figure~\ref{fig:lineprof}. For $^{12}$CO $J=2-1$, we deproject and azimuthally average the velocity integrated line flux. The $1\sigma$ error bar on the line ratio or line flux, including a 15\% calibration uncertainty combined in quadrature with the observed RMS on the individual observations is provided in the right panel for each constraint. We only plot the model and data where the corresponding lines are detected at $>3\sigma$ (or would have been detected, in the case of the models). Note that the shape of the curve is better constrained (i.e. not affected by the 15\% calibration error) than the vertical offset. For $^{12}$CO and $^{13}$CO the shape is regulated mainly by the CO reduction factor. C$^{18}$O ratios depend slightly more on $G_0$. The high $G_0$ models ($\ge 16$) are flatter due to warmer gas temperatures at large radii. 


\section{Discussion}\label{sec:discussion}

Using resolved observations of multiple isotopologues and rotational transitions of CO, millimeter continuum, the SED,  and detailed chemical modeling, we have constructed a new global model for the IM Lup protoplanetary disk structure. This model is generally consistent with previous work from \citet{pinte2008} regarding the continuum and \citet{panic2009} for the distribution of gas. With the high quality ALMA data presented here, we can constrain the distributions of gas and dust, the gas temperatures, the CO abundances, the CO optical depths, and the incident external radiation field. In addition, our detailed models for CO provide predictive power for other molecular species based upon the full set of abundances computed by the chemical code.

\subsection{Gas versus dust concentration}
We confirm the earlier results of \citet{panic2009} that the gas and small grains are significantly more extended than the large grains. At the resolution of our observations, the gas distribution appears relatively smooth and extends out to 970~AU. The resolved millimeter continuum suggests the large grain distribution has two components, a bright central region that is $\le40$~AU in diameter and a broad halo with a shallow power-law slope of $\Sigma\propto R^{-0.3}$, truncated at $R=313$~AU. This extended dust component traces a substantial amount of millimeter-sized dust at large radii. Why these grains have not continued to drift inward remains to be seen, although there are hints of substructure (see Section~\ref{sec:rings}) that may help to alleviate the problem.

\subsection{Gas mass and accretion}
The total dust mass from both large and small grains in our model is $M_{\rm dust} = 1.7\times10^{-3}$~$M_\odot$, from which we derive a gas mass of $M_{\rm gas} = 0.17$~$M_\odot$ for a gas to dust mass ratio of 100. This large mass is somewhat surprising given that this source has been previously characterized as a low accretor. \citet{padgett2006} identified it as a borderline weak-line T Tauri / classical T Tauri star based on its variable H-$\alpha$ equivalent width in 2004. From data taken in 2008 -- 2009, \citet{gunther2010} suggested a mass accretion rate of $\dot{M}\le10^{-11}$~${M}_{\odot}$~year$^{-1}$, based on the absence of veiled photospheric emission along with narrow H-$\alpha$. Earlier measurements demonstrate significant H-$\alpha$ variability \citep{herbig1988,batalha1993,batalha1998,wichmann1999}, even on day to day timescales, but with a small line equivalent width. More recently, \citet{salyk2013} provided an accretion luminosity estimate for IM Lup based on the Pf$\beta$ luminosity. Correcting for the adopted distance between that work and ours and using our stellar mass and radius, the mass accretion rate is more typical of a classical T Tauri, $9\times10^{-9}$~${M}_{\odot}$~year$^{-1}$; however, with the scatter in the accretion luminosity versus Pf$\beta$ luminosity relationship, the possible range spans $3.6\times10^{-10} - 2.5\times10^{-7}$~${{M}}_{\odot}$~year$^{-1}$.  It is possible that we are catching IM Lup between accretion outbursts, where the bright inner disk seen in the Band 6 and 7 continuum could be related to piled up and/or hot material. Longer term monitoring of H-$\alpha$ and other accretion tracers coupled with ALMA continuum imaging can further explore this scenario as a potential explanation for the low accretion rate yet high disk mass.

\subsection{CO abundance and isotopologues}
Assuming the dust-derived gas mass, we find that CO is under-abundant by a factor of 20 based on all CO data, including $^{12}$CO. Given the young age of the system \citep[$<1$~Myr;][]{mawet2012}, this finding points to early chemical depletion of carbon \citep[e.g.,][]{bergin2014,furuya2014,yu2016}, or depletion in a previous evolutionary phase. The CO deficit cannot be explained by selective photodissociation alone, since the finding is supported by $^{12}$CO as well as CO isotopologue modeling. We do, however, find that the C$^{18}$O models favor a factor of $\gtrsim2$ times more CO reduction than the $^{12}$CO or $^{13}$CO data, which is consistent with models that include isotope selective photodissociation \citep{miotello2014}. As a result, the IM Lup disk may be an excellent test case for these types of isotopic studies, along with fractionation studies for other molecules. 

Alternatively, if the CO abundance is more typical and the disk is instead gas-depleted, the disk gas mass would be a factor of 20 lower or $M_{\rm gas} = 9\times10^{-3}$~$M_\odot$, i.e., a gas to dust mass ratio of 5. The dust mass is far less uncertain. Even including uncertainties arising from the unknown nature of the central unresolved component, the dust and pebble mass is known within 25\%. If we do not include the unresolved inner disk given its high optical depth, the gas to dust mass ratio would be 6.25. 

Our chemical modeling also provides insight into the various CO formation pathways in this disk. \citet{oberg2015im} discovered a ring of DCO$^+$ emission located outside of the millimeter continuum edge in the IM Lup disk. Our models predict an increase in both midplane CO abundance and CO column density at this same location due to photon-driven processes. Therefore the DCO$^+$ ring could naturally arise from this CO enhancement. Unlike other disks,  including AS 209 \citep{huang2016} and TW Hya \citep{schwarz2016}, IM Lup's CO enhancement is not directly visible in the CO isotopologue data. Perhaps the change in integrated column density ($\sim20\%$) is not sufficient, or countered by the change in gas-weighted temperature (e.g., Figure~\ref{fig:mod_column}). Deeper, higher resolution observations of CO isotopologues IM Lup will need to be made to see if this CO enhancement indeed is present.

\subsection{Constraints on the local interstellar radiation field}
Another interesting aspect of the models are the constraints on the external radiation field from the CO observations, where we find $G_0 \le 4$ based on the global line fluxes and the line ratio profiles. The Lupus star-forming complex is relatively low mass, similar to the Taurus star-forming region though with perhaps a larger fraction of lower mass stars \citep{hughes1994}. However, nearby star-forming regions, including Upper Scorpius, Upper Centaurus Lupus, and Lower Centaurus Crux have substantially more massive stellar populations. Each of these regions hosts at least $\sim40$ B-type stars, though between them only one O-type star \citep[$\zeta$~Oph, an O9V star;][]{mamajek2002,preibisch2008}. As a complementary estimate for $G_0$, we have taken the known O, B, and A stars within 1~kpc from the {\em Hipparcos} catalogue and estimated the external FUV from each at IM~Lup's location. We adopt the latest DR1 {\it Gaia} distance for the source of 161~pc. These calculations are described in Appendix~\ref{app:hipp}. The {\it Hipparcos} estimated external $G_0$ is between $2.9-4.5$, which is consistent with our estimate from the CO observations. From these calculations, cross-listed with the young O and B star catalogue of \citet{dezeeuw1999}, we find that that the external UV originates from $\sim60\%$ field O and B stars, and $\sim40\%$ known young B-type stars in neighboring clusters (primarily Upper Centaurus Lupus).

\subsection{Inner Disk Dust Opacity}\label{sec:opa}
While our models reproduce the presence of an inner drop in the CO isotopologues emission (Figure~\ref{fig:mom0}), the depth of the depletion in our models is less pronounced than what is observed. Correspondingly, our models may not be fully capturing the magnitude of the dust optical depth towards the center of the disk. Based on our models, which are first constrained by the dust SED and the 875~$\mu$m visibilities, the dust optical depth in the inner $R\le20$~AU disk is $\tau_{875\mu}\ge4$, and $\tau_{875\mu}>10$ at 1~AU. Beyond this radius, the models have $\tau_{875\mu}\lesssim1$, and so should be more closely tracing mass. As a result, there may be dust in the inner disk that our observations are not sensitive to. Longer wavelength, e.g., ALMA Band 3 or centimenter-wavelength observations, would help better clarify the nature of the inner disk. 

As a test of the effects of additional dust, we have recomputed the line intensity of our standard CO abundance models for the $^{13}$CO $J=3-2$ transition with $10\times$ enhanced dust mass inside of 50~AU ($0\farcs31$), just larger than our resolution limit and show the results in Figure~\ref{fig:innerdust}. The drop in the inner $^{13}$CO $J=3-2$ line flux is strongly sensitive to the amount of dust in the inner disk, where dust absorption is blocking a significant fraction of the emergent line flux. Going forward, gas observations of disk systems with highly concentrated dust may be strongly hampered by dust opacity, even in the $J=2-1$ observations in ALMA Band 6. In a gas/dust survey of the Lupus star forming region with ALMA, \citet{ansdell2016} present at least four disks where a ring in $^{13}$CO is seen where there is centrally peaked continuum. This scenario is directly opposite from what is observed in transition disks, where the gas is more centrally concentrated than the millimeter dust gap \citep[e.g.,][]{vandermarel2015}.

Alternatively, our opacity model could be too simplistic. There have been previous studies demonstrating how temperature can increase the dust absorption cross section for a given dust composition \citep{agladze1996,mennella1998}. Given that the inner disk -- the warmest region -- is the one most strongly affected, this is certainly a possible explanation. For example, \citet{mennella1998} shows that the absorption opacity of amorphous iron-rich olivine (FAYA) increases by a factor of $\sim3$ with a temperature increase from 20~K to 100~K. These temperature (and composition) dependent effects should be included in future, more detailed, models.

\begin{figure}[t]
\begin{centering}
\includegraphics[width=0.41\textwidth]{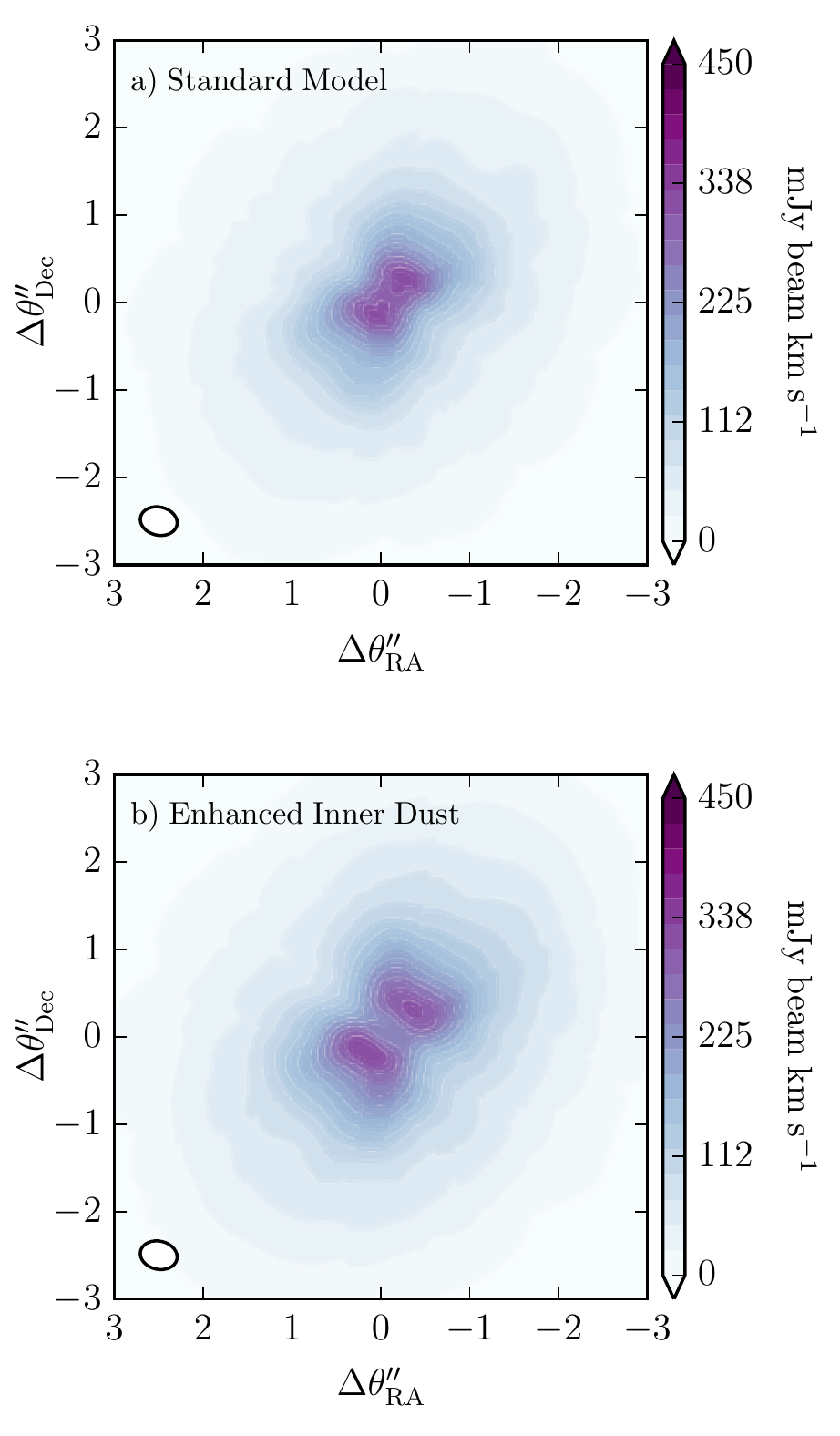}
\caption{a) Model moment 0 map for for $^{13}$CO $J=3-2$. The simulated hole is shallower than the observations (Figure~\ref{fig:mom0}). Given that the inner disk continuum is already optically thick, we have tested models with $10\times$ more dust inside of 50~AU ($0\farcs31$) shown in panel b). The excess dust reduces the observed CO in the inner disk substantially and reduces the CO within the hole to within the observed range.  \label{fig:innerdust}}
\end{centering}
\end{figure}

\subsection{Tentative Dust Rings}\label{sec:rings}
Axisymmetric narrow rings in the millimeter continuum have been observed in the Class I HL~Tau protostellar disk \citep{alma2015} and the Class II TW Hya disk \citep{andrews2016}, and may be a relatively common phenomenon based on oscillating structure in the real visibilities of other bright disks \citep{zhang2016}. While our data have lower resolution than the previous ring detections ($0\farcs3$ or $\sim50$~AU), there are intriguing breaks present in the observed deprojected Band 7 continuum brightness profile near $0\farcs95$ (150~AU) and $1\farcs55$ (250~AU). The variations in the continuum brightness profile are quite small, deviating by $\sim2-3\%$ from a smooth profile. The discontinuity is more clearly visible in the slope versus radius (see Figure~\ref{fig:dustprof}, top). Applying an unsharp mask to the continuum data and subtracting off 90\% of the smoothed flux reveals ring-like structures, and highlights their axisymmetry (Figure~\ref{fig:dustprof}, bottom).   Higher resolution observations are necessary to confirm the presence of these rings and understand their structure. We do not expect shallow, narrow rings to substantially change the bulk density and temperature properties of the model, especially for the gas; however, wide and deep rings may allow more stellar radiation to reach and heat the disk midplane.
\begin{figure}[t]
\begin{centering}
\includegraphics[width=0.39\textwidth]{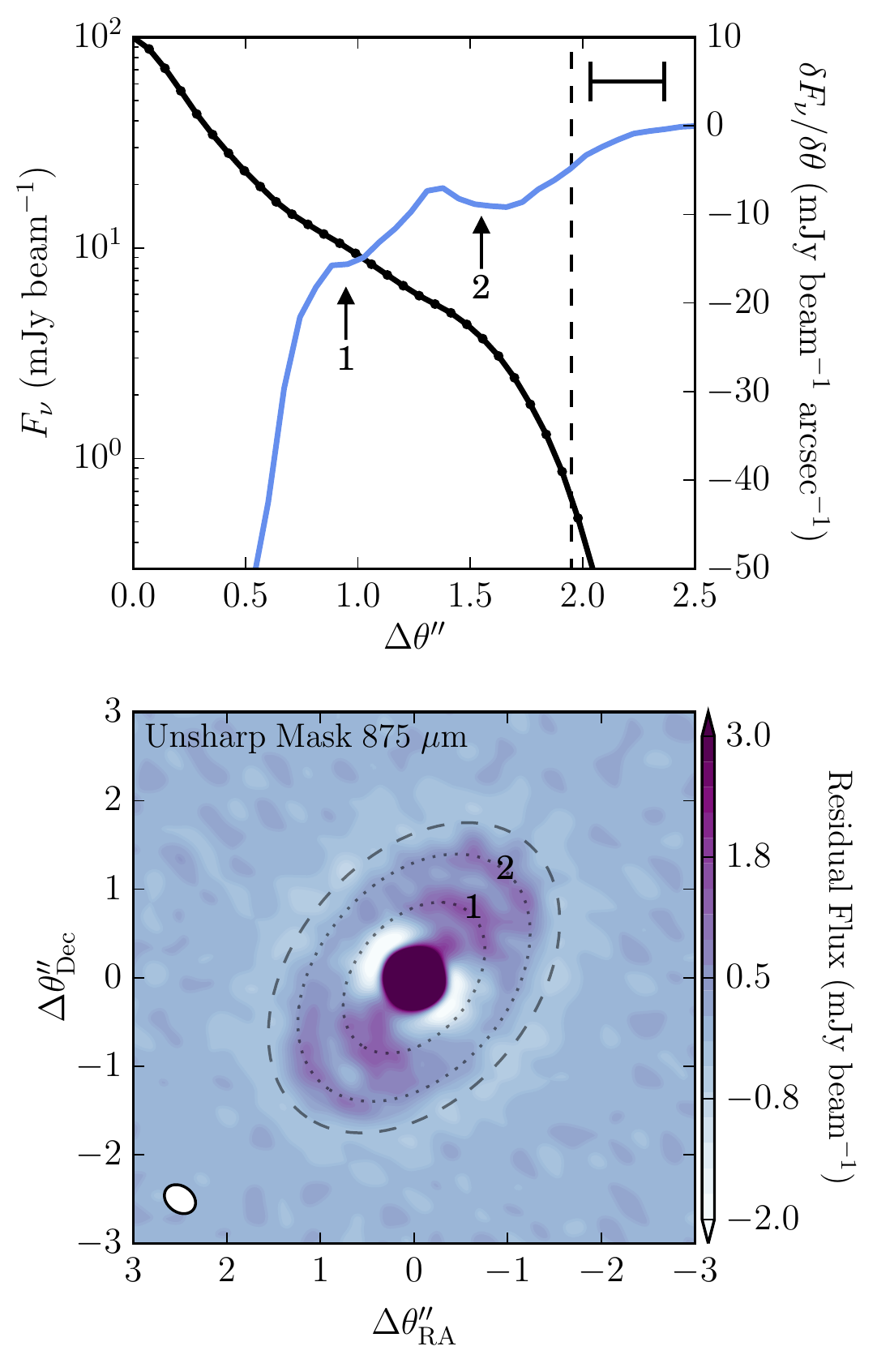}
\caption{Deprojected radial continuum at 875$\mu$m. The continuum has shallow breaks in the slope (spatial resolution indicated by the horizontal bar at the top right). These features are even more clearly seen in the continuum slope versus radius shown in blue. Ring locations at 1) $0\farcs95$ ($\sim150$~AU) and 2) $1\farcs55$ ($\sim250$~AU) are indicated by arrows (top) and dotted ellipses (bottom). The outer disk radius is indicated by the long-dash line. Note the color scale is centrally saturated to highlight the ring features.  \label{fig:dustprof}}
\end{centering}
\end{figure}

\subsection{Diffuse $^{12}$CO}\label{sec:diffuse}
There are two interesting features in the central channels of $^{12}$CO $J=2-1$ that our models do not reproduce. First, at velocities $\lesssim1$~km~s$^{-1}$, the inner disk is very faint compared to higher velocity channels. Figure~\ref{fig:diffuse} shows the velocity averaged channel maps of $^{12}$CO $J=2-1$ compared to the best-fit model. While excess dust in the center helps decrease the visible line flux, it affects the isotopologues more strongly than $^{12}$CO. Second, a diffuse halo of $^{12}$CO extended emission spreads out laterally away from the star to large radii ($\sim970$~AU). The contribution from the extended diffuse emission is also seen in the deprojected, azimuthally averaged integrated line profiles from Figure~\ref{fig:lineprof}, top row, which includes these central channels. The best fit model of $f_{\rm CO}=0.05$ and $G_0\le4$ fits the observed profile of $^{12}$CO $J=2-1$ reasonably well out to $\sim3\farcs5$ (560~AU), but beyond this radius fails to reproduce the observed shallow $^{12}$CO plateau out to $\sim6''$ (820 AU).  
\begin{figure*} 
\begin{centering}
\includegraphics[width=1.0\textwidth]{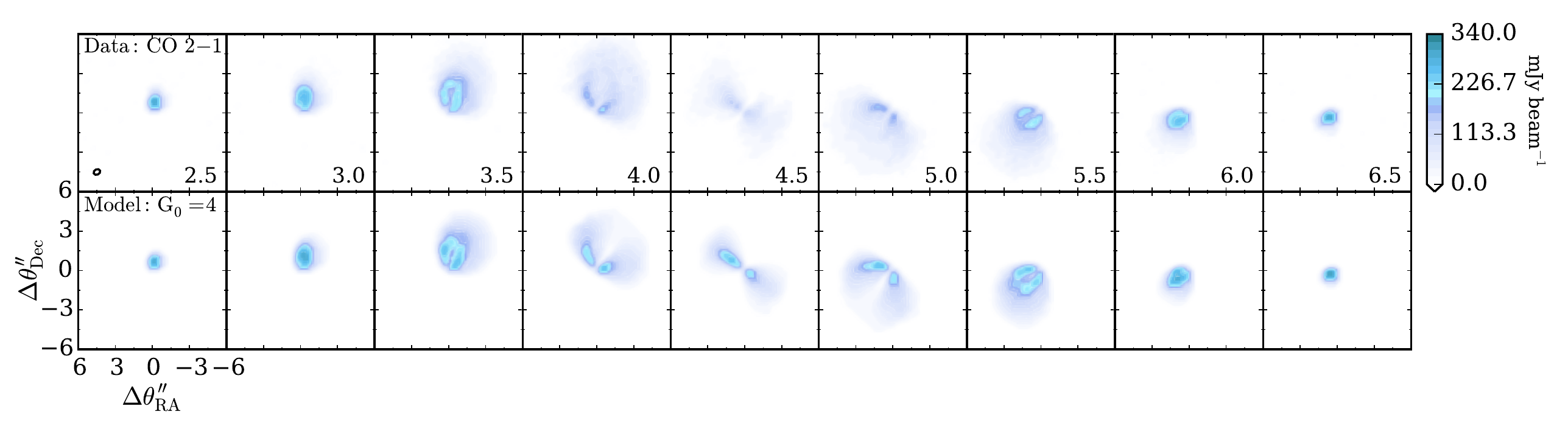}
\caption{Low velocity CO ``halo'' in the central channels of $^{12}$CO $J=2-1$ as compared to the model CO distributions for $f_{\rm CO}=0.05$. The central channels contain a substantial amount of diffuse CO at large distances from the source that is slightly asymmetric, with fainter than expected $^{12}$CO emission near the center. \label{fig:diffuse}}
\end{centering}
\end{figure*}

There are a number of possible explanations for the peculiar low velocity $^{12}$CO $J=2-1$ morphology:
\begin{enumerate}
\item An incorrect model surface density.
\item Turbulent velocity broadening.
\item Photoevaporative winds.
\item Foreground cloud contamination.
\item A bound envelope (natal remnant or gravitationally accumulated).
\end{enumerate}
The first three potential explanations are related to incorrect assumptions in our underlying model, while the latter two describe additional model components that are not included within the present framework. Below we discuss each of these scenarios further.

The gas density structure could simply be incorrect, i.e., scenario 1). The disk gas may extend to larger radii than the model or may have an entirely different surface density profile. Our models reasonably match the CO isotopologue emission, suggesting that CO at high densities ($n_{\rm H_2}>10^6$~cm$^{-3}$) is well reproduced; however, the CO present in the lower density atmosphere is not well constrained.  
For scenario 2), we tested models with twice the turbulence (doppler broadening), 200~m~s$^{-1}$, and find that they do not produce enough emission at large radii, and also ``puff out'' the emission toward the star in the $\pm0.5$~km~s$^{-1}$ channels into a bean-shaped structure, which is not what is observed. Rather the $^{12}$CO appears broadened only radially outward, away from the star in these channels. The preferential outward-broadening could be alternatively explained by a photoevaporative wind, i.e., scenario 3). Models of photoevaporative winds are typically much hotter than what our models suggest \citep[e.g.,][]{clarke2001,alexander2006,gorti2009,owen2012}, and the external $G_0$ is relatively low ($\le4$) to launch an externally driven wind \citep{facchini2016}. 

In scenario 4), IM Lup is known to be associated with foreground cloud material, which could provide a source of $^{12}$CO absorption; however, the cloud velocity is offset from the source \citep{vankempen2007,panic2009}. Finally, the diffuse emission could trace a tenuous, gravitationally bound remnant envelope. \citet{pinte2008} noted a halo of scattered light that extended beyond IM Lup's dark lane (taken to be the location of the disk), which they attributed to an envelope. A more spherical structure, such as a vertically puffed up torus, as opposed to a flattened disk could explain the wide angle diffuse emission. Furthermore, if this material is cold ($<20$~K) and present at large radii ($\gtrsim1000$~AU) with a wide covering fraction (i.e., more spherical than flattened in nature) it could provide a source of foreground absorption for the low velocity $^{12}$CO $J=2-1$, making the central channels appear fainter everywhere.

To test the final envelope/torus absorption scenario, we can estimate the foreground CO column required to attenuate the observed emission. The model predicts a peak brightness of 253~mJy~beam$^{-1}$ in the central channel, compared to the observed 147~mJy~beam$^{-1}$. This reduction corresponds to a line center optical depth of 0.54. For an envelope temperature of 10~K and line width of 1.5~km~s$^{-1}$, the total CO column density required is $N_{\rm CO}=5.4\times10^{16}$~cm$^{-2}$. Using the standard interstellar CO abundance, this corresponds to a visual extinction of just $\tau_{\rm V}=0.3$, consistent with \citet{pinte2008}'s estimates for their suggested dusty envelope.

We cannot tell the difference between a remnant envelope and gas gravitationally captured. Since the system is young, $\le 1$~Myr \citep{mawet2012}, a remnant envelope would not be surprising. The FWHM of the diffuse emission is about $\sim1$~km~s$^{-1}$, which corresponds to a radial size scale of $\sim2200$~AU. Based on the estimates above, its average density would be $n_{\rm H} = 8\times10^3$~cm$^{-3}$, consistent with the \citet{whitney2003b} models of a natal envelope of a late Class I/early Class II object. 

To estimate whether the second envelope scenario -- gas capture -- is feasible, we can compare this size scale to IM Lup's Bondi radius \citep{bondi1952}. The turbulent velocity in Lupus~2 is about $\Delta v \sim 0.34-0.79$~km~s$^{-1}$ \citep{hara1999}. Taking the upper value implies IM Lup's region of influence is about $r_B\sim3000$~AU. The observed structure is well within this scale. Regardless, these scales are going to be strongly spatially filtered by ALMA, and thus would require lower frequency or total power observations to fully characterize the nature of this kind of extended component.

\section{Summary}\label{sec:summary}
We present a new model for the combined gas and dust structure of the IM Lup protoplanetary disk. The millimeter grains have a two component distribution with a bright inner disk and a broad halo that is truncated sharply at 313~AU (see Figure~\ref{fig:mom0}). The gas disk extends out to at least 570~AU and perhaps as far as 970~AU, where the outer regions may form an extension of the disk or be associated with a gravitationally bound envelope. As was seen by \citet{pinte2008}, the disk is quite flared and massive, $M_{\rm gas}=0.17$~$M_\odot$, assuming an interstellar gas to dust ratio, which may be a result of its youth \citep[$\le1$~Myr;][]{mawet2012}. 

We find that the dust in the inner disk is sufficiently optically thick to hide line emission. The less abundant isotopologues are especially affected, creating an inner emission deficit where the disk continuum is brightest. The depth of the emission deficit ($>2$) suggests that the millimeter continuum is becoming optically thick above the midplane, implying that there are large grains vertically suspended in the inner disk. Perhaps this young system has not had time to settle, turbulence is lofting grains efficiently, or magnetic suspension is maintaining grains high up in the disk atmosphere \citep[e.g.,][]{turner2014}. Higher resolution observations will help shed light on this bright inner disk component that is just below our resolution.

The analysis presented in this paper demonstrates the utility of multiple lines of different isotopologues of CO combined with sensitive continuum constraints for understanding the coupled gas and dust structure of protoplanetary disks. Both the absolute line fluxes and ratios help shed light on the nature of the CO gas, its abundance, and the disk temperature structure. Several puzzles remain, however, which require new observations to resolve. In summary, we find:
\begin{enumerate}
\item A massive yet gravitationally stable ($Q>1$) disk with $M_{\rm gas} = 0.17~M_\odot$ around a 1~$M_\odot$ star. The gas is $\gtrsim2\times$ more radially extended than the millimeter grains.
\item A chemically derived external radiation field of $G_0\le4$, which matches a {\it Hipparcos} census of nearby massive stars. Because of Lupus' low mass stellar population, field stars contribute more than half of the external $G_0$ of $2.9-4.5$.
\item An optically thick inner $\lesssim40$~AU disk in the Band 6 and 7 continuum, at least in part causing the inner flux deficit seen in the CO isotopologues.
\item Minor breaks in the continuum slope suggestive of ring-like structures.
\item An extended diffuse halo of $^{12}$CO related to an extension of the disk or an extended envelope, associated with the extended scattered light reported in \citet{pinte2008}.
\end{enumerate}

\acknowledgements{{\it Acknowledgements:} The authors thank Michiel Hogerheijde, James Owen, and Richard Teague for useful discussions, along with the anonymous referee. This paper makes use of the following ALMA data: ADS/JAO.ALMA\#2013.00694 and ADS/JAO.ALMA\#2013.1.00226.S. ALMA is a partnership of ESO (representing its member states), NSF (USA) and NINS (Japan), together with NRC (Canada) and NSC and ASIAA (Taiwan), in cooperation with the Republic of Chile. The Joint ALMA Observatory is operated by ESO, AUI/NRAO and NAOJ. The National Radio Astronomy Observatory is a facility of the National Science Foundation operated under cooperative agreement by Associated Universities, Inc. We acknowledge the use of public data from the Swift data archive.
LIC acknowledges the support of NASA through Hubble Fellowship grant HST-HF2-51356.001-A awarded by the Space Telescope Science Institute, which is operated by the Association of Universities for Research in Astronomy, Inc., for NASA, under contract NAS 5-26555. 
KI\"O also acknowledges funding through a Packard Fellowship for Science and Engineering from the David and Lucile Packard Foundation.
JH and RAL acknowledge support by the National Science Foundation Graduate Research Fellowship under Grant No. DGE-1144152. IC gratefully acknowledges funding support from the Smithsonian Institution.
}


\begin{thebibliography}{117}
\expandafter\ifx\csname natexlab\endcsname\relax\def\natexlab#1{#1}\fi

\bibitem[{{Adams}(2010)}]{adams2010}
{Adams}, F.~C. 2010, \araa, 48, 47

\bibitem[{{Agladze} {et~al.}(1996){Agladze}, {Sievers}, {Jones}, {Burlitch}, \&
  {Beckwith}}]{agladze1996}
{Agladze}, N.~I., {Sievers}, A.~J., {Jones}, S.~A., {Burlitch}, J.~M., \&
  {Beckwith}, S.~V.~W. 1996, \apj, 462, 1026

\bibitem[{{Alencar} \& {Batalha}(2002)}]{alencar2002}
{Alencar}, S.~H.~P. \& {Batalha}, C. 2002, \apj, 571, 378

\bibitem[{{Alexander} {et~al.}(2006){Alexander}, {Clarke}, \&
  {Pringle}}]{alexander2006}
{Alexander}, R.~D., {Clarke}, C.~J., \& {Pringle}, J.~E. 2006, \mnras, 369, 229

\bibitem[{{ALMA Partnership} {et~al.}(2015){ALMA Partnership}, {Brogan},
  {P{\'e}rez}, {Hunter}, {Dent}, {Hales}, {Hills}, {Corder}, {Fomalont},
  {Vlahakis}, {Asaki}, {Barkats}, {Hirota}, {Hodge}, {Impellizzeri}, {Kneissl},
  {Liuzzo}, {Lucas}, {Marcelino}, {Matsushita}, {Nakanishi}, {Phillips},
  {Richards}, {Toledo}, {Aladro}, {Broguiere}, {Cortes}, {Cortes}, {Espada},
  {Galarza}, {Garcia-Appadoo}, {Guzman-Ramirez}, {Humphreys}, {Jung}, {Kameno},
  {Laing}, {Leon}, {Marconi}, {Mignano}, {Nikolic}, {Nyman}, {Radiszcz},
  {Remijan}, {Rod{\'o}n}, {Sawada}, {Takahashi}, {Tilanus}, {Vila Vilaro},
  {Watson}, {Wiklind}, {Akiyama}, {Chapillon}, {de Gregorio-Monsalvo}, {Di
  Francesco}, {Gueth}, {Kawamura}, {Lee}, {Nguyen Luong}, {Mangum}, {Pietu},
  {Sanhueza}, {Saigo}, {Takakuwa}, {Ubach}, {van Kempen}, {Wootten},
  {Castro-Carrizo}, {Francke}, {Gallardo}, {Garcia}, {Gonzalez}, {Hill},
  {Kaminski}, {Kurono}, {Liu}, {Lopez}, {Morales}, {Plarre}, {Schieven},
  {Testi}, {Videla}, {Villard}, {Andreani}, {Hibbard}, \&
  {Tatematsu}}]{alma2015}
{ALMA Partnership}, {Brogan}, C.~L., {P{\'e}rez}, L.~M., {Hunter}, T.~R.,
  {Dent}, W.~R.~F., {Hales}, A.~S., {Hills}, R.~E., {Corder}, S., {Fomalont},
  E.~B., {Vlahakis}, C., {Asaki}, Y., {Barkats}, D., {Hirota}, A., {Hodge},
  J.~A., {Impellizzeri}, C.~M.~V., {Kneissl}, R., {Liuzzo}, E., {Lucas}, R.,
  {Marcelino}, N., {Matsushita}, S., {Nakanishi}, K., {Phillips}, N.,
  {Richards}, A.~M.~S., {Toledo}, I., {Aladro}, R., {Broguiere}, D., {Cortes},
  J.~R., {Cortes}, P.~C., {Espada}, D., {Galarza}, F., {Garcia-Appadoo}, D.,
  {Guzman-Ramirez}, L., {Humphreys}, E.~M., {Jung}, T., {Kameno}, S., {Laing},
  R.~A., {Leon}, S., {Marconi}, G., {Mignano}, A., {Nikolic}, B., {Nyman},
  L.-A., {Radiszcz}, M., {Remijan}, A., {Rod{\'o}n}, J.~A., {Sawada}, T.,
  {Takahashi}, S., {Tilanus}, R.~P.~J., {Vila Vilaro}, B., {Watson}, L.~C.,
  {Wiklind}, T., {Akiyama}, E., {Chapillon}, E., {de Gregorio-Monsalvo}, I.,
  {Di Francesco}, J., {Gueth}, F., {Kawamura}, A., {Lee}, C.-F., {Nguyen
  Luong}, Q., {Mangum}, J., {Pietu}, V., {Sanhueza}, P., {Saigo}, K.,
  {Takakuwa}, S., {Ubach}, C., {van Kempen}, T., {Wootten}, A.,
  {Castro-Carrizo}, A., {Francke}, H., {Gallardo}, J., {Garcia}, J.,
  {Gonzalez}, S., {Hill}, T., {Kaminski}, T., {Kurono}, Y., {Liu}, H.-Y.,
  {Lopez}, C., {Morales}, F., {Plarre}, K., {Schieven}, G., {Testi}, L.,
  {Videla}, L., {Villard}, E., {Andreani}, P., {Hibbard}, J.~E., \&
  {Tatematsu}, K. 2015, \apjl, 808, L3

\bibitem[{{Andrews} {et~al.}(2011){Andrews}, {Wilner}, {Espaillat}, {Hughes},
  {Dullemond}, {McClure}, {Qi}, \& {Brown}}]{andrews2011}
{Andrews}, S.~M., {Wilner}, D.~J., {Espaillat}, C., {Hughes}, A.~M.,
  {Dullemond}, C.~P., {McClure}, M.~K., {Qi}, C., \& {Brown}, J.~M. 2011, \apj,
  732, 42

\bibitem[{{Andrews} {et~al.}(2012){Andrews}, {Wilner}, {Hughes}, {Qi},
  {Rosenfeld}, {{\"O}berg}, {Birnstiel}, {Espaillat}, {Cieza}, {Williams},
  {Lin}, \& {Ho}}]{andrews2012}
{Andrews}, S.~M., {Wilner}, D.~J., {Hughes}, A.~M., {Qi}, C., {Rosenfeld},
  K.~A., {{\"O}berg}, K.~I., {Birnstiel}, T., {Espaillat}, C., {Cieza}, L.~A.,
  {Williams}, J.~P., {Lin}, S.-Y., \& {Ho}, P.~T.~P. 2012, \apj, 744, 162

\bibitem[{{Andrews} {et~al.}(2016){Andrews}, {Wilner}, {Zhu}, {Birnstiel},
  {Carpenter}, {P{\'e}rez}, {Bai}, {{\"O}berg}, {Hughes}, {Isella}, \&
  {Ricci}}]{andrews2016}
{Andrews}, S.~M., {Wilner}, D.~J., {Zhu}, Z., {Birnstiel}, T., {Carpenter},
  J.~M., {P{\'e}rez}, L.~M., {Bai}, X.-N., {{\"O}berg}, K.~I., {Hughes}, A.~M.,
  {Isella}, A., \& {Ricci}, L. 2016, \apjl, 820, L40

\bibitem[{{Ansdell} {et~al.}(2016){Ansdell}, {Williams}, {van der Marel},
  {Carpenter}, {Guidi}, {Hogerheijde}, {Mathews}, {Manara}, {Miotello},
  {Natta}, {Oliveira}, {Tazzari}, {Testi}, {van Dishoeck}, \& {van
  Terwisga}}]{ansdell2016}
{Ansdell}, M., {Williams}, J.~P., {van der Marel}, N., {Carpenter}, J.~M.,
  {Guidi}, G., {Hogerheijde}, M., {Mathews}, G.~S., {Manara}, C.~F.,
  {Miotello}, A., {Natta}, A., {Oliveira}, I., {Tazzari}, M., {Testi}, L., {van
  Dishoeck}, E.~F., \& {van Terwisga}, S.~E. 2016, ArXiv e-prints

\bibitem[{{Bai} \& {Goodman}(2009)}]{bai2009}
{Bai}, X.-N. \& {Goodman}, J. 2009, \apj, 701, 737

\bibitem[{{Batalha} \& {Basri}(1993)}]{batalha1993}
{Batalha}, C.~C. \& {Basri}, G. 1993, \apj, 412, 363

\bibitem[{{Batalha} {et~al.}(1998){Batalha}, {Quast}, {Torres}, {Pereira},
  {Terra}, {Jablonski}, {Schiavon}, {de La Reza}, \& {Sartori}}]{batalha1998}
{Batalha}, C.~C., {Quast}, G.~R., {Torres}, C.~A.~O., {Pereira}, P.~C.~R.,
  {Terra}, M.~A.~O., {Jablonski}, F., {Schiavon}, R.~P., {de La Reza}, J.~R.,
  \& {Sartori}, M.~J. 1998, \aaps, 128, 561

\bibitem[{{Bergin}(2013)}]{bergin2013}
{Bergin}, E.~A. 2013, ArXiv e-prints

\bibitem[{{Bergin} {et~al.}(2014){Bergin}, {Cleeves}, {Crockett}, \&
  {Blake}}]{bergin2014}
{Bergin}, E.~A., {Cleeves}, L.~I., {Crockett}, N., \& {Blake}, G.~A. 2014,
  Faraday Discussions, 168, 61

\bibitem[{{Bergin} {et~al.}(2013){Bergin}, {Cleeves}, {Gorti}, {Zhang},
  {Blake}, {Green}, {Andrews}, {Evans}, {Henning}, {{\"O}berg}, {Pontoppidan},
  {Qi}, {Salyk}, \& {van Dishoeck}}]{bergin2013hd}
{Bergin}, E.~A., {Cleeves}, L.~I., {Gorti}, U., {Zhang}, K., {Blake}, G.~A.,
  {Green}, J.~D., {Andrews}, S.~M., {Evans}, II, N.~J., {Henning}, T.,
  {{\"O}berg}, K., {Pontoppidan}, K., {Qi}, C., {Salyk}, C., \& {van Dishoeck},
  E.~F. 2013, \nat, 493, 644

\bibitem[{{Bethell} \& {Bergin}(2011)}]{bethell2011u}
{Bethell}, T.~J. \& {Bergin}, E.~A. 2011, \apj, 739, 78

\bibitem[{{Birnstiel} \& {Andrews}(2014)}]{birnstiel2014}
{Birnstiel}, T. \& {Andrews}, S.~M. 2014, \apj, 780, 153

\bibitem[{{Bohlin} {et~al.}(1978){Bohlin}, {Savage}, \& {Drake}}]{bohlin1978}
{Bohlin}, R.~C., {Savage}, B.~D., \& {Drake}, J.~F. 1978, \apj, 224, 132

\bibitem[{{Bondi}(1952)}]{bondi1952}
{Bondi}, H. 1952, \mnras, 112, 195

\bibitem[{{Brinch} \& {Hogerheijde}(2010)}]{brinch2010}
{Brinch}, C. \& {Hogerheijde}, M.~R. 2010, \aap, 523, A25

\bibitem[{{Bruderer}(2013)}]{bruderer2013}
{Bruderer}, S. 2013, \aap, 559, A46

\bibitem[{{Clarke} {et~al.}(2001){Clarke}, {Gendrin}, \&
  {Sotomayor}}]{clarke2001}
{Clarke}, C.~J., {Gendrin}, A., \& {Sotomayor}, M. 2001, \mnras, 328, 485

\bibitem[{{Cleeves}(2016)}]{cleeves2016a}
{Cleeves}, L.~I. 2016, \apjl, 816, L21

\bibitem[{{Cleeves} {et~al.}(2013){Cleeves}, {Adams}, \&
  {Bergin}}]{cleeves2013a}
{Cleeves}, L.~I., {Adams}, F.~C., \& {Bergin}, E.~A. 2013, \apj, 772, 5

\bibitem[{{Cleeves} {et~al.}(2014){Cleeves}, {Bergin}, \&
  {Adams}}]{cleeves2014par}
{Cleeves}, L.~I., {Bergin}, E.~A., \& {Adams}, F.~C. 2014, \apj, 794, 123

\bibitem[{{Cleeves} {et~al.}(2015){Cleeves}, {Bergin}, {Qi}, {Adams}, \&
  {{\"O}berg}}]{cleeves2015tw}
{Cleeves}, L.~I., {Bergin}, E.~A., {Qi}, C., {Adams}, F.~C., \& {{\"O}berg},
  K.~I. 2015, \apj, 799, 204

\bibitem[{{Comer{\'o}n}(2008)}]{comeron2008}
{Comer{\'o}n}, F. {The Lupus Clouds}, ed. B.~{Reipurth}, 295

\bibitem[{{Covino} {et~al.}(1992){Covino}, {Terranegra}, {Franchini},
  {Chavarria-K.}, \& {Stalio}}]{covino1992}
{Covino}, E., {Terranegra}, L., {Franchini}, M., {Chavarria-K.}, C., \&
  {Stalio}, R. 1992, \aaps, 94, 273

\bibitem[{{Cuzzi} {et~al.}(1993){Cuzzi}, {Dobrovolskis}, \&
  {Champney}}]{cuzzi1993}
{Cuzzi}, J.~N., {Dobrovolskis}, A.~R., \& {Champney}, J.~M. 1993, \icarus, 106,
  102

\bibitem[{{de Gregorio-Monsalvo} {et~al.}(2013){de Gregorio-Monsalvo},
  {M{\'e}nard}, {Dent}, {Pinte}, {L{\'o}pez}, {Klaassen}, {Hales},
  {Cort{\'e}s}, {Rawlings}, {Tachihara}, {Testi}, {Takahashi}, {Chapillon},
  {Mathews}, {Juhasz}, {Akiyama}, {Higuchi}, {Saito}, {Nyman}, {Phillips},
  {Rod{\'o}n}, {Corder}, \& {Van Kempen}}]{degregorio2013}
{de Gregorio-Monsalvo}, I., {M{\'e}nard}, F., {Dent}, W., {Pinte}, C.,
  {L{\'o}pez}, C., {Klaassen}, P., {Hales}, A., {Cort{\'e}s}, P., {Rawlings},
  M.~G., {Tachihara}, K., {Testi}, L., {Takahashi}, S., {Chapillon}, E.,
  {Mathews}, G., {Juhasz}, A., {Akiyama}, E., {Higuchi}, A.~E., {Saito}, M.,
  {Nyman}, L.-{\AA}., {Phillips}, N., {Rod{\'o}n}, J., {Corder}, S., \& {Van
  Kempen}, T. 2013, \aap, 557, A133

\bibitem[{{de Zeeuw} {et~al.}(1999){de Zeeuw}, {Hoogerwerf}, {de Bruijne},
  {Brown}, \& {Blaauw}}]{dezeeuw1999}
{de Zeeuw}, P.~T., {Hoogerwerf}, R., {de Bruijne}, J.~H.~J., {Brown}, A.~G.~A.,
  \& {Blaauw}, A. 1999, \aj, 117, 354

\bibitem[{{Draine} \& {Lee}(1984)}]{draine1984}
{Draine}, B.~T. \& {Lee}, H.~M. 1984, \apj, 285, 89

\bibitem[{{Dullemond} \& {Dominik}(2004)}]{dullemond2004}
{Dullemond}, C.~P. \& {Dominik}, C. 2004, \aap, 421, 1075

\bibitem[{{ESA}(1997)}]{hipparcos1997}
{ESA}, ed. 1997, ESA Special Publication, Vol. 1200, {The HIPPARCOS and TYCHO
  catalogues. Astrometric and photometric star catalogues derived from the ESA
  HIPPARCOS Space Astrometry Mission}

\bibitem[{{Facchini} {et~al.}(2016){Facchini}, {Clarke}, \&
  {Bisbas}}]{facchini2016}
{Facchini}, S., {Clarke}, C.~J., \& {Bisbas}, T.~G. 2016, \mnras, 457, 3593

\bibitem[{{Favre} {et~al.}(2013){Favre}, {Cleeves}, {Bergin}, {Qi}, \&
  {Blake}}]{favre2013}
{Favre}, C., {Cleeves}, L.~I., {Bergin}, E.~A., {Qi}, C., \& {Blake}, G.~A.
  2013, \apjl, 776, L38

\bibitem[{{Fock} {et~al.}(1980){Fock}, {G{\"u}rtler}, \& {Koch}}]{fock1980}
{Fock}, J.-H., {G{\"u}rtler}, P., \& {Koch}, E.~E. 1980, Chemical Physics, 47,
  87

\bibitem[{{Fogel} {et~al.}(2011){Fogel}, {Bethell}, {Bergin}, {Calvet}, \&
  {Semenov}}]{fogel2011}
{Fogel}, J.~K.~J., {Bethell}, T.~J., {Bergin}, E.~A., {Calvet}, N., \&
  {Semenov}, D. 2011, \apj, 726, 29

\bibitem[{{Furlan} {et~al.}(2006){Furlan}, {Hartmann}, {Calvet}, {D'Alessio},
  {Franco-Hern{\'a}ndez}, {Forrest}, {Watson}, {Uchida}, {Sargent}, {Green},
  {Keller}, \& {Herter}}]{furlan2006}
{Furlan}, E., {Hartmann}, L., {Calvet}, N., {D'Alessio}, P.,
  {Franco-Hern{\'a}ndez}, R., {Forrest}, W.~J., {Watson}, D.~M., {Uchida},
  K.~I., {Sargent}, B., {Green}, J.~D., {Keller}, L.~D., \& {Herter}, T.~L.
  2006, \apjs, 165, 568

\bibitem[{{Furuya} \& {Aikawa}(2014)}]{furuya2014}
{Furuya}, K. \& {Aikawa}, Y. 2014, \apj, 790, 97

\bibitem[{{Gaia Collaboration} {et~al.}(2016){Gaia Collaboration}, {Brown},
  {Vallenari}, {Prusti}, {de Bruijne}, {Mignard}, {Drimmel}, \&
  {co-authors}}]{gaia}
{Gaia Collaboration}, {Brown}, A.~G.~A., {Vallenari}, A., {Prusti}, T., {de
  Bruijne}, J., {Mignard}, F., {Drimmel}, R., \& {co-authors}, . 2016, ArXiv
  e-prints

\bibitem[{{Galli} {et~al.}(2013){Galli}, {Bertout}, {Teixeira}, \&
  {Ducourant}}]{galli2013}
{Galli}, P.~A.~B., {Bertout}, C., {Teixeira}, R., \& {Ducourant}, C. 2013,
  \aap, 558, A77

\bibitem[{{Goldreich} \& {Ward}(1973)}]{goldreich1973}
{Goldreich}, P. \& {Ward}, W.~R. 1973, \apj, 183, 1051

\bibitem[{{Gorti} \& {Hollenbach}(2009)}]{gorti2009}
{Gorti}, U. \& {Hollenbach}, D. 2009, \apj, 690, 1539

\bibitem[{{Gorti} {et~al.}(2015){Gorti}, {Hollenbach}, \&
  {Dullemond}}]{gorti2015}
{Gorti}, U., {Hollenbach}, D., \& {Dullemond}, C.~P. 2015, \apj, 804, 29

\bibitem[{{Gottlieb} \& {Upson}(1969)}]{gottlieb1969}
{Gottlieb}, D.~M. \& {Upson}, II, W.~L. 1969, \apj, 157, 611

\bibitem[{{Guilloteau} {et~al.}(2016){Guilloteau}, {Pi{\'e}tu}, {Chapillon},
  {Di Folco}, {Dutrey}, {Henning}, {Semenov}, {Birnstiel}, \&
  {Grosso}}]{guilloteau2016}
{Guilloteau}, S., {Pi{\'e}tu}, V., {Chapillon}, E., {Di Folco}, E., {Dutrey},
  A., {Henning}, T., {Semenov}, D., {Birnstiel}, T., \& {Grosso}, N. 2016,
  \aap, 586, L1

\bibitem[{{G{\"u}nther} {et~al.}(2010){G{\"u}nther}, {Matt}, {Schmitt},
  {G{\"u}del}, {Li}, \& {Burton}}]{gunther2010}
{G{\"u}nther}, H.~M., {Matt}, S.~P., {Schmitt}, J.~H.~M.~M., {G{\"u}del}, M.,
  {Li}, Z.-Y., \& {Burton}, D.~M. 2010, \aap, 519, A97

\bibitem[{{Habing}(1968)}]{habing}
{Habing}, H.~J. 1968, \bain, 19, 421

\bibitem[{{Hara} {et~al.}(1999){Hara}, {Tachihara}, {Mizuno}, {Onishi},
  {Kawamura}, {Obayashi}, \& {Fukui}}]{hara1999}
{Hara}, A., {Tachihara}, K., {Mizuno}, A., {Onishi}, T., {Kawamura}, A.,
  {Obayashi}, A., \& {Fukui}, Y. 1999, \pasj, 51, 895

\bibitem[{{Harries}(2000)}]{harries2000}
{Harries}, T.~J. 2000, \mnras, 315, 722

\bibitem[{{Harries} {et~al.}(2004){Harries}, {Monnier}, {Symington}, \&
  {Kurosawa}}]{harries2004}
{Harries}, T.~J., {Monnier}, J.~D., {Symington}, N.~H., \& {Kurosawa}, R. 2004,
  \mnras, 350, 565

\bibitem[{{Henkel} {et~al.}(1994){Henkel}, {Wilson}, {Langer}, {Chin}, \&
  {Mauersberger}}]{henkel1994}
{Henkel}, C., {Wilson}, T.~L., {Langer}, N., {Chin}, Y.-N., \& {Mauersberger},
  R. 1994, in Lecture Notes in Physics, Berlin Springer Verlag, Vol. 439, The
  Structure and Content of Molecular Clouds, ed. T.~L. {Wilson} \& K.~J.
  {Johnston}, 72--88

\bibitem[{{Herbig} \& {Bell}(1988)}]{herbig1988}
{Herbig}, G.~H. \& {Bell}, K.~R. 1988, {Third Catalog of Emission-Line Stars of
  the Orion Population : 3 : 1988}

\bibitem[{{Herczeg} {et~al.}(2002){Herczeg}, {Linsky}, {Valenti},
  {Johns-Krull}, \& {Wood}}]{herczeg2002}
{Herczeg}, G.~J., {Linsky}, J.~L., {Valenti}, J.~A., {Johns-Krull}, C.~M., \&
  {Wood}, B.~E. 2002, \apj, 572, 310

\bibitem[{{Herczeg} {et~al.}(2004){Herczeg}, {Wood}, {Linsky}, {Valenti}, \&
  {Johns-Krull}}]{herczeg2004}
{Herczeg}, G.~J., {Wood}, B.~E., {Linsky}, J.~L., {Valenti}, J.~A., \&
  {Johns-Krull}, C.~M. 2004, \apj, 607, 369

\bibitem[{{Hogerheijde} {et~al.}(2016){Hogerheijde}, {Bekkers}, {Pinilla},
  {Salinas}, {Kama}, {Andrews}, {Qi}, \& {Wilner}}]{hogerheijde2016}
{Hogerheijde}, M.~R., {Bekkers}, D., {Pinilla}, P., {Salinas}, V.~N., {Kama},
  M., {Andrews}, S.~M., {Qi}, C., \& {Wilner}, D.~J. 2016, \aap, 586, A99

\bibitem[{{Huang} {et~al.}(2016){Huang}, {{\"O}berg}, \& {Andrews}}]{huang2016}
{Huang}, J., {{\"O}berg}, K.~I., \& {Andrews}, S.~M. 2016, \apjl, 823, L18

\bibitem[{{Hughes} {et~al.}(2011){Hughes}, {Wilner}, {Andrews}, {Qi}, \&
  {Hogerheijde}}]{hughes2011}
{Hughes}, A.~M., {Wilner}, D.~J., {Andrews}, S.~M., {Qi}, C., \& {Hogerheijde},
  M.~R. 2011, \apj, 727, 85

\bibitem[{{Hughes} {et~al.}(1994){Hughes}, {Hartigan}, {Krautter}, \&
  {Kelemen}}]{hughes1994}
{Hughes}, J., {Hartigan}, P., {Krautter}, J., \& {Kelemen}, J. 1994, \aj, 108,
  1071

\bibitem[{{Ingleby} {et~al.}(2013){Ingleby}, {Calvet}, {Herczeg}, {Blaty},
  {Walter}, {Ardila}, {Alexander}, {Edwards}, {Espaillat}, {Gregory},
  {Hillenbrand}, \& {Brown}}]{ingleby2013}
{Ingleby}, L., {Calvet}, N., {Herczeg}, G., {Blaty}, A., {Walter}, F.,
  {Ardila}, D., {Alexander}, R., {Edwards}, S., {Espaillat}, C., {Gregory},
  S.~G., {Hillenbrand}, L., \& {Brown}, A. 2013, \apj, 767, 112

\bibitem[{{Isella} {et~al.}(2007){Isella}, {Testi}, {Natta}, {Neri}, {Wilner},
  \& {Qi}}]{isella2007}
{Isella}, A., {Testi}, L., {Natta}, A., {Neri}, R., {Wilner}, D., \& {Qi}, C.
  2007, \aap, 469, 213

\bibitem[{{Jankowski} \& {Szalewicz}(2005)}]{jankowski2005}
{Jankowski}, P. \& {Szalewicz}, K. 2005, \jcp, 123, 104301

\bibitem[{{Johnson}(1966)}]{johnson1966}
{Johnson}, H.~L. 1966, \araa, 4, 193

\bibitem[{Kama {et~al.}(2016)Kama, Bruderer, Carney, Hogerheijde, van Dishoeck,
  Fedele, Baryshev, Boland, G{\"u}sten, Aikutalp, Choi, Endo, Frieswijk,
  Karska, Klaassen, Koumpia, Kristensen, Leurini, Nagy, Beaupuits, Risacher,
  van~der Marel, van Kempen, van Weeren, Wyrowski, \& Y{\i}ld{\i}z}]{kama2016}
Kama, M., Bruderer, S., Carney, M., Hogerheijde, M., van Dishoeck, E.~F.,
  Fedele, D., Baryshev, A., Boland, W., G{\"u}sten, R., Aikutalp, A., Choi, Y.,
  Endo, A., Frieswijk, W., Karska, A., Klaassen, P., Koumpia, E., Kristensen,
  L., Leurini, S., Nagy, Z., Beaupuits, J. P.~P., Risacher, C., van~der Marel,
  N., van Kempen, T.~A., van Weeren, R.~J., Wyrowski, F., \& Y{\i}ld{\i}z,
  U.~A. 2016

\bibitem[{{Kurosawa} {et~al.}(2004){Kurosawa}, {Harries}, {Bate}, \&
  {Symington}}]{kurosawa2004}
{Kurosawa}, R., {Harries}, T.~J., {Bate}, M.~R., \& {Symington}, N.~H. 2004,
  \mnras, 351, 1134

\bibitem[{{Kurucz}(1979)}]{kurucz1979}
{Kurucz}, R.~L. 1979, \apjs, 40, 1

\bibitem[{{Lombardi} {et~al.}(2008){Lombardi}, {Lada}, \&
  {Alves}}]{lombardi2008}
{Lombardi}, M., {Lada}, C.~J., \& {Alves}, J. 2008, \aap, 480, 785

\bibitem[{{Lommen} {et~al.}(2007){Lommen}, {Wright}, {Maddison},
  {J{\o}rgensen}, {Bourke}, {van Dishoeck}, {Hughes}, {Wilner}, {Burton}, \&
  {van Langevelde}}]{lommen2007}
{Lommen}, D., {Wright}, C.~M., {Maddison}, S.~T., {J{\o}rgensen}, J.~K.,
  {Bourke}, T.~L., {van Dishoeck}, E.~F., {Hughes}, A., {Wilner}, D.~J.,
  {Burton}, M., \& {van Langevelde}, H.~J. 2007, \aap, 462, 211

\bibitem[{{Lommen} {et~al.}(2010){Lommen}, {van Dishoeck}, {Wright},
  {Maddison}, {Min}, {Wilner}, {Salter}, {van Langevelde}, {Bourke}, {van der
  Burg}, \& {Blake}}]{lommen2010}
{Lommen}, D.~J.~P., {van Dishoeck}, E.~F., {Wright}, C.~M., {Maddison}, S.~T.,
  {Min}, M., {Wilner}, D.~J., {Salter}, D.~M., {van Langevelde}, H.~J.,
  {Bourke}, T.~L., {van der Burg}, R.~F.~J., \& {Blake}, G.~A. 2010, \aap, 515,
  A77

\bibitem[{{Lucy}(1999)}]{lucy1999}
{Lucy}, L.~B. 1999, \aap, 344, 282

\bibitem[{{Lynden-Bell} \& {Pringle}(1974)}]{lyndenbell1974}
{Lynden-Bell}, D. \& {Pringle}, J.~E. 1974, \mnras, 168, 603

\bibitem[{{Mamajek} {et~al.}(2002){Mamajek}, {Meyer}, \&
  {Liebert}}]{mamajek2002}
{Mamajek}, E.~E., {Meyer}, M.~R., \& {Liebert}, J. 2002, \aj, 124, 1670

\bibitem[{{Mathis}(1990)}]{mathis1990}
{Mathis}, J.~S. 1990, \araa, 28, 37

\bibitem[{{Mathis} {et~al.}(1977){Mathis}, {Rumpl}, \& {Nordsieck}}]{mrn1977}
{Mathis}, J.~S., {Rumpl}, W., \& {Nordsieck}, K.~H. 1977, \apj, 217, 425

\bibitem[{{Mawet} {et~al.}(2012){Mawet}, {Absil}, {Montagnier}, {Riaud},
  {Surdej}, {Ducourant}, {Augereau}, {R{\"o}ttinger}, {Girard}, {Krist}, \&
  {Stapelfeldt}}]{mawet2012}
{Mawet}, D., {Absil}, O., {Montagnier}, G., {Riaud}, P., {Surdej}, J.,
  {Ducourant}, C., {Augereau}, J.-C., {R{\"o}ttinger}, S., {Girard}, J.,
  {Krist}, J., \& {Stapelfeldt}, K. 2012, \aap, 544, A131

\bibitem[{{McClure} {et~al.}(2016){McClure}, {Bergin}, {Cleeves}, {van
  Dishoeck}, {Blake}, {Evans}, {Green}, {Henning}, {{\"O}berg}, {Pontoppidan},
  \& {Salyk}}]{mcclure2016}
{McClure}, M., {Bergin}, T., {Cleeves}, I., {van Dishoeck}, E., {Blake}, G.,
  {Evans}, N., {Green}, J., {Henning}, T., {{\"O}berg}, K., {Pontoppidan}, K.,
  \& {Salyk}, C. 2016, ArXiv e-prints

\bibitem[{{McMullin} {et~al.}(2007){McMullin}, {Waters}, {Schiebel}, {Young},
  \& {Golap}}]{mcmullin2007}
{McMullin}, J.~P., {Waters}, B., {Schiebel}, D., {Young}, W., \& {Golap}, K.
  2007, in Astronomical Society of the Pacific Conference Series, Vol. 376,
  Astronomical Data Analysis Software and Systems XVI, ed. R.~A. {Shaw},
  F.~{Hill}, \& D.~J. {Bell}, 127

\bibitem[{{Mennella} {et~al.}(1998){Mennella}, {Brucato}, {Colangeli},
  {Palumbo}, {Rotundi}, \& {Bussoletti}}]{mennella1998}
{Mennella}, V., {Brucato}, J.~R., {Colangeli}, L., {Palumbo}, P., {Rotundi},
  A., \& {Bussoletti}, E. 1998, \apj, 496, 1058

\bibitem[{{Miotello} {et~al.}(2014){Miotello}, {Bruderer}, \& {van
  Dishoeck}}]{miotello2014}
{Miotello}, A., {Bruderer}, S., \& {van Dishoeck}, E.~F. 2014, \aap, 572, A96

\bibitem[{{{\"O}berg} {et~al.}(2011){{\"O}berg}, {Boogert}, {Pontoppidan}, {van
  den Broek}, {van Dishoeck}, {Bottinelli}, {Blake}, \&
  {Evans}}]{oberg2011spitz}
{{\"O}berg}, K.~I., {Boogert}, A.~C.~A., {Pontoppidan}, K.~M., {van den Broek},
  S., {van Dishoeck}, E.~F., {Bottinelli}, S., {Blake}, G.~A., \& {Evans}, II,
  N.~J. 2011, \apj, 740, 109

\bibitem[{{{\"O}berg} {et~al.}(2015){{\"O}berg}, {Furuya}, {Loomis}, {Aikawa},
  {Andrews}, {Qi}, {van Dishoeck}, \& {Wilner}}]{oberg2015im}
{{\"O}berg}, K.~I., {Furuya}, K., {Loomis}, R., {Aikawa}, Y., {Andrews}, S.~M.,
  {Qi}, C., {van Dishoeck}, E.~F., \& {Wilner}, D.~J. 2015, \apj, 810, 112

\bibitem[{{Owen} {et~al.}(2012){Owen}, {Clarke}, \& {Ercolano}}]{owen2012}
{Owen}, J.~E., {Clarke}, C.~J., \& {Ercolano}, B. 2012, \mnras, 422, 1880

\bibitem[{{Padgett} {et~al.}(2006){Padgett}, {Cieza}, {Stapelfeldt}, {Evans},
  {Koerner}, {Sargent}, {Fukagawa}, {van Dishoeck}, {Augereau}, {Allen},
  {Blake}, {Brooke}, {Chapman}, {Harvey}, {Porras}, {Lai}, {Mundy}, {Myers},
  {Spiesman}, \& {Wahhaj}}]{padgett2006}
{Padgett}, D.~L., {Cieza}, L., {Stapelfeldt}, K.~R., {Evans}, II, N.~J.,
  {Koerner}, D., {Sargent}, A., {Fukagawa}, M., {van Dishoeck}, E.~F.,
  {Augereau}, J.-C., {Allen}, L., {Blake}, G., {Brooke}, T., {Chapman}, N.,
  {Harvey}, P., {Porras}, A., {Lai}, S.-P., {Mundy}, L., {Myers}, P.~C.,
  {Spiesman}, W., \& {Wahhaj}, Z. 2006, \apj, 645, 1283

\bibitem[{{Pani{\'c}} {et~al.}(2009){Pani{\'c}}, {Hogerheijde}, {Wilner}, \&
  {Qi}}]{panic2009}
{Pani{\'c}}, O., {Hogerheijde}, M.~R., {Wilner}, D., \& {Qi}, C. 2009, \aap,
  501, 269

\bibitem[{{Pecaut} \& {Mamajek}(2013)}]{pecaut2013}
{Pecaut}, M.~J. \& {Mamajek}, E.~E. 2013, \apjs, 208, 9

\bibitem[{{Pecaut} {et~al.}(2012){Pecaut}, {Mamajek}, \& {Bubar}}]{pecaut2012}
{Pecaut}, M.~J., {Mamajek}, E.~E., \& {Bubar}, E.~J. 2012, \apj, 746, 154

\bibitem[{{Pi{\'e}tu} {et~al.}(2007){Pi{\'e}tu}, {Dutrey}, \&
  {Guilloteau}}]{pietu2007}
{Pi{\'e}tu}, V., {Dutrey}, A., \& {Guilloteau}, S. 2007, \aap, 467, 163

\bibitem[{{Pinte} {et~al.}(2016){Pinte}, {Dent}, {M{\'e}nard}, {Hales}, {Hill},
  {Cortes}, \& {de Gregorio-Monsalvo}}]{pinte2016}
{Pinte}, C., {Dent}, W.~R.~F., {M{\'e}nard}, F., {Hales}, A., {Hill}, T.,
  {Cortes}, P., \& {de Gregorio-Monsalvo}, I. 2016, \apj, 816, 25

\bibitem[{{Pinte} {et~al.}(2009){Pinte}, {Harries}, {Min}, {Watson},
  {Dullemond}, {Woitke}, {M{\'e}nard}, \& {Dur{\'a}n-Rojas}}]{pinte2009}
{Pinte}, C., {Harries}, T.~J., {Min}, M., {Watson}, A.~M., {Dullemond}, C.~P.,
  {Woitke}, P., {M{\'e}nard}, F., \& {Dur{\'a}n-Rojas}, M.~C. 2009, \aap, 498,
  967

\bibitem[{{Pinte} \& {Laibe}(2014)}]{pinte2014}
{Pinte}, C. \& {Laibe}, G. 2014, \aap, 565, A129

\bibitem[{{Pinte} {et~al.}(2008){Pinte}, {Padgett}, {M{\'e}nard},
  {Stapelfeldt}, {Schneider}, {Olofsson}, {Pani{\'c}}, {Augereau},
  {Duch{\^e}ne}, {Krist}, {Pontoppidan}, {Perrin}, {Grady}, {Kessler-Silacci},
  {van Dishoeck}, {Lommen}, {Silverstone}, {Hines}, {Wolf}, {Blake}, {Henning},
  \& {Stecklum}}]{pinte2008}
{Pinte}, C., {Padgett}, D.~L., {M{\'e}nard}, F., {Stapelfeldt}, K.~R.,
  {Schneider}, G., {Olofsson}, J., {Pani{\'c}}, O., {Augereau}, J.~C.,
  {Duch{\^e}ne}, G., {Krist}, J., {Pontoppidan}, K., {Perrin}, M.~D., {Grady},
  C.~A., {Kessler-Silacci}, J., {van Dishoeck}, E.~F., {Lommen}, D.,
  {Silverstone}, M., {Hines}, D.~C., {Wolf}, S., {Blake}, G.~A., {Henning}, T.,
  \& {Stecklum}, B. 2008, \aap, 489, 633

\bibitem[{{Prantzos} {et~al.}(1996){Prantzos}, {Aubert}, \&
  {Audouze}}]{prantzos1996}
{Prantzos}, N., {Aubert}, O., \& {Audouze}, J. 1996, \aap, 309, 760

\bibitem[{{Preibisch} \& {Mamajek}(2008)}]{preibisch2008}
{Preibisch}, T. \& {Mamajek}, E. {The Nearest OB Association:
  Scorpius-Centaurus (Sco OB2)}, ed. B.~{Reipurth}, 235

\bibitem[{Qi {et~al.}(2011)Qi, D'alessio, {\"O}berg, Wilner, Hughes, Andrews,
  \& Ayala}]{qi2011}
Qi, C., D'alessio, P., {\"O}berg, K.~I., Wilner, D.~J., Hughes, A.~M., Andrews,
  S.~M., \& Ayala, S. 2011

\bibitem[{{Rosenfeld} {et~al.}(2013{\natexlab{a}}){Rosenfeld}, {Andrews},
  {Hughes}, {Wilner}, \& {Qi}}]{rosenfeld2013b}
{Rosenfeld}, K.~A., {Andrews}, S.~M., {Hughes}, A.~M., {Wilner}, D.~J., \&
  {Qi}, C. 2013{\natexlab{a}}, \apj, 774, 16

\bibitem[{{Rosenfeld} {et~al.}(2013{\natexlab{b}}){Rosenfeld}, {Andrews},
  {Wilner}, {Kastner}, \& {McClure}}]{rosenfeld2013a}
{Rosenfeld}, K.~A., {Andrews}, S.~M., {Wilner}, D.~J., {Kastner}, J.~H., \&
  {McClure}, M.~K. 2013{\natexlab{b}}, \apj, 775, 136

\bibitem[{{Salyk} {et~al.}(2013){Salyk}, {Herczeg}, {Brown}, {Blake},
  {Pontoppidan}, \& {van Dishoeck}}]{salyk2013}
{Salyk}, C., {Herczeg}, G.~J., {Brown}, J.~M., {Blake}, G.~A., {Pontoppidan},
  K.~M., \& {van Dishoeck}, E.~F. 2013, \apj, 769, 21

\bibitem[{{Sch{\"o}ier} {et~al.}(2005){Sch{\"o}ier}, {van der Tak}, {van
  Dishoeck}, \& {Black}}]{schoier2005}
{Sch{\"o}ier}, F.~L., {van der Tak}, F.~F.~S., {van Dishoeck}, E.~F., \&
  {Black}, J.~H. 2005, \aap, 432, 369

\bibitem[{{Schwarz} {et~al.}(2016){Schwarz}, {Bergin}, {Cleeves}, {Blake},
  {Zhang}, {{\"O}berg}, {van Dishoeck}, \& {Qi}}]{schwarz2016}
{Schwarz}, K.~R., {Bergin}, E.~A., {Cleeves}, L.~I., {Blake}, G.~A., {Zhang},
  K., {{\"O}berg}, K.~I., {van Dishoeck}, E.~F., \& {Qi}, C. 2016, \apj, 823,
  91

\bibitem[{{Smith} {et~al.}(2004){Smith}, {Herbst}, \& {Chang}}]{smith2004}
{Smith}, I.~W.~M., {Herbst}, E., \& {Chang}, Q. 2004, \mnras, 350, 323

\bibitem[{{Teague} {et~al.}(2016){Teague}, {Guilloteau}, {Semenov}, {Henning},
  {Dutrey}, {Pietu}, {Birnstiel}, {Chapillon}, {Hollenbach}, \&
  {Gorti}}]{teague2016}
{Teague}, R., {Guilloteau}, S., {Semenov}, D., {Henning}, T., {Dutrey}, A.,
  {Pietu}, V., {Birnstiel}, T., {Chapillon}, E., {Hollenbach}, D., \& {Gorti},
  U. 2016, ArXiv e-prints

\bibitem[{{Trumpler}(1930)}]{trumpler1930}
{Trumpler}, R.~J. 1930, \pasp, 42, 214

\bibitem[{{Turner} {et~al.}(2014){Turner}, {Benisty}, {Dullemond}, \&
  {Hirose}}]{turner2014}
{Turner}, N.~J., {Benisty}, M., {Dullemond}, C.~P., \& {Hirose}, S. 2014, \apj,
  780, 42

\bibitem[{{van der Marel} {et~al.}(2015){van der Marel}, {van Dishoeck},
  {Bruderer}, {P{\'e}rez}, \& {Isella}}]{vandermarel2015}
{van der Marel}, N., {van Dishoeck}, E.~F., {Bruderer}, S., {P{\'e}rez}, L., \&
  {Isella}, A. 2015, \aap, 579, A106

\bibitem[{{van Kempen} {et~al.}(2007){van Kempen}, {van Dishoeck}, {Brinch}, \&
  {Hogerheijde}}]{vankempen2007}
{van Kempen}, T.~A., {van Dishoeck}, E.~F., {Brinch}, C., \& {Hogerheijde},
  M.~R. 2007, \aap, 461, 983

\bibitem[{{van Leeuwen}(2007)}]{leeuwen2007}
{van Leeuwen}, F. 2007, \aap, 474, 653

\bibitem[{{Walsh} {et~al.}(2014){Walsh}, {Juh{\'a}sz}, {Pinilla}, {Harsono},
  {Mathews}, {Dent}, {Hogerheijde}, {Birnstiel}, {Meeus}, {Nomura}, {Aikawa},
  {Millar}, \& {Sandell}}]{walsh2014}
{Walsh}, C., {Juh{\'a}sz}, A., {Pinilla}, P., {Harsono}, D., {Mathews}, G.~S.,
  {Dent}, W.~R.~F., {Hogerheijde}, M.~R., {Birnstiel}, T., {Meeus}, G.,
  {Nomura}, H., {Aikawa}, Y., {Millar}, T.~J., \& {Sandell}, G. 2014, \apjl,
  791, L6

\bibitem[{{Weidenschilling}(1977)}]{weidenschilling1977}
{Weidenschilling}, S.~J. 1977, \apss, 51, 153

\bibitem[{{Weidenschilling}(1980)}]{weidenschilling1980}
---. 1980, \icarus, 44, 172

\bibitem[{{Whipple}(1972)}]{whipple1972}
{Whipple}, F.~L. 1972, in From Plasma to Planet, ed. A.~{Elvius}, 211

\bibitem[{{Whitney} {et~al.}(2003){Whitney}, {Wood}, {Bjorkman}, \&
  {Cohen}}]{whitney2003b}
{Whitney}, B.~A., {Wood}, K., {Bjorkman}, J.~E., \& {Cohen}, M. 2003, \apj,
  598, 1079

\bibitem[{{Wichmann} {et~al.}(1999){Wichmann}, {Covino}, {Alcal{\'a}},
  {Krautter}, {Allain}, \& {Hauschildt}}]{wichmann1999}
{Wichmann}, R., {Covino}, E., {Alcal{\'a}}, J.~M., {Krautter}, J., {Allain},
  S., \& {Hauschildt}, P.~H. 1999, \mnras, 307, 909

\bibitem[{{Wright} {et~al.}(2003){Wright}, {Egan}, {Kraemer}, \&
  {Price}}]{wright2003}
{Wright}, C.~O., {Egan}, M.~P., {Kraemer}, K.~E., \& {Price}, S.~D. 2003, \aj,
  125, 359

\bibitem[{{Yang} {et~al.}(2010){Yang}, {Stancil}, {Balakrishnan}, \&
  {Forrey}}]{yang2010}
{Yang}, B., {Stancil}, P.~C., {Balakrishnan}, N., \& {Forrey}, R.~C. 2010,
  \apj, 718, 1062

\bibitem[{{Yu} {et~al.}(2016){Yu}, {Willacy}, {Dodson-Robinson}, {Turner}, \&
  {Evans}}]{yu2016}
{Yu}, M., {Willacy}, K., {Dodson-Robinson}, S.~E., {Turner}, N.~J., \& {Evans},
  II, N.~J. 2016, \apj, 822, 53

\bibitem[{{Zhang} {et~al.}(2016){Zhang}, {Bergin}, {Blake}, {Cleeves},
  {Hogerheijde}, {Salinas}, \& {Schwarz}}]{zhang2016}
{Zhang}, K., {Bergin}, E.~A., {Blake}, G.~A., {Cleeves}, L.~I., {Hogerheijde},
  M., {Salinas}, V., \& {Schwarz}, K.~R. 2016, \apjl, 818, L16

\end{thebibliography}

\appendix
\section{External Radiation Field Estimated from {\it Hipparcos}}\label{app:hipp}

In this section we describe a secondary technique to estimate the external UV field at IM Lup's position. Using the astrometric data provided in the {\it Hipparcos} database \citep{hipparcos1997}, we have compiled catalogue data for stars satisfying spectral types of [OBA*] and parallax $\ge1.0$~mas ($d_\odot\le1$~kpc). From this initial database, we have updated the spectral types where available from the \citet{wright2003} Tycho-2 spectral type catalogue and parallaxes using the new reduction of \citet{leeuwen2007}. The final catalogue contains 23,315 local O, B and A-type stars.

For each star, we estimate their individual UV flux contribution to the IM Lup disk using their 3D galactic positions (parallax plus on sky position) with the astropy {\texttt{coordinates}} package\footnote{\url{http://docs.astropy.org/en/stable/coordinates/}}. Because the inter-object line of sight extinction is not well constrained, we use constant values for the UV extinction per kpc, which we vary between no extinction ($A_{\rm UV}~d^{-1}=0$ mag kpc$^{-1}$) and higher than galactic average extinction ($A_{\rm UV}~d^{-1}=3$ mag kpc$^{-1}$, with an average value of $A_{\rm UV}=3A_{\rm V} = 2.1$ per kpc). Given that recent star formation activity has cleared out large volumes of the local ISM in the region \citep[see, e.g., reviews of][]{comeron2008,preibisch2008}, there will be significant extinction variations depending on the particular line of sight, and as such the ``true'' answer is likely in between these values. We also assume for simplicity that the stars emit as blackbodies at their catalogue identified effective temperature.  To test this assumption we have compared stellar Kurucz models for main sequence dwarfs \citep{kurucz1979} against the blackbody prediction and find that for an O8V (B0V) star a blackbody underestimates the flux by $\sim30\%$ ($25\%$). Thus we expect the unknown line of sight extinction to be the dominant source of uncertainty in these approximate calculations.

For the known dwarf stars, we adopt the stellar radius from the table kindly provided online\footnote{\url{http://www.pas.rochester.edu/~emamajek/EEM_dwarf_UBVIJHK_colors_Teff.txt}} by Erik Mamajek \citep{pecaut2012,pecaut2013} to estimate the flux at IM Lup, i.e., $F(\nu) = B_*(\nu, T_{\rm eff}) \left(R_* / R_{\rm IML}\right)^2$. For the giants, we calculate an approximate stellar radius based on the parallax distance and the observed Johnson V-band magnitude, adding a crude extinction correction of $A_{\rm V} = 0.7$~mag~pc$^{-1}$ between us and the star \citep{trumpler1930,gottlieb1969}. The corrected V-band magnitude is compared to the star's blackbody estimate integrated over the Johnson V-band transmission function and converted to flux using the tables provided in \citet{johnson1966}.
With the observed and ``intrinsic'' fluxes, combined with the parallax distance, we are able to estimate a stellar radius.
 
Based on the radius derived from the Mamajek table for the dwarfs and $V_J$ for the giants/others, we can calculate the integrated FUV flux at each star's surface from $912-2800$~\AA\ and then estimate its contribution to IM Lup using their 3D galactic positions. This wavelength range is considered as it is relevant for heating via the photoelectric effect. CO photodissociation will only be impacted by line-processes between $912-1118$~\AA\ \citep{fock1980} and in recomputing $G_0$ using this range we find that the $G_0$ relevant for photodissociation is $\sim4\times$ less than the value relevant for heating. 

For stars with missing luminosity classes, complicated identifiers, or missing temperatures (592 sources, 2.5\%), we followed the same procedure as for the giants, but took the temperature from the Mamajek table based on the star's temperature class. If there was no temperature subclass provided in the updated \citet{leeuwen2007} spectral type, we did not include the flux in our calculation (38 stars, 0.16\%). 

From the total catalogue, we identified O and B type stars associated with star-forming regions using the tables provided in \citet{dezeeuw1999}. All other stars are classified as ``field stars'' though some  unclassified young stars may be included in this distinction. Table~\ref{tab:g0} lists the calculated $G_0$ from young stars and field stars for different extinction assumptions.  For a distance of 161~pc, field stars provide $\sim60\%$ of the external radiation while near-field young B-stars, largely associated with Upper Centaurus Lupus, provide about $\sim40\%$. 
Figure~\ref{fig:externalG0} illustrates the stellar distribution for the top 50 stars contributing to IM Lup's external $G_0$, totaling $\sim80\%$ of the total $G_0= 3.3$ for an interstellar UV extinction of 2~mag~kpc$^{-1}$ and $d_{\rm IML}=161$~pc. Early B-type stars (B0 -- B2) close $\lesssim100$~pc to IM Lup are the dominant contributor with a few field O-type stars (the earliest being O6V, HD 42088) showing up as well even though they are at $\gtrsim100$~pc distances. We find that late B-type and early A-type stars do not contribute significantly to $G_0$.

\begin{figure*}[bh!]
\begin{centering}
\includegraphics[width=1.0\textwidth]{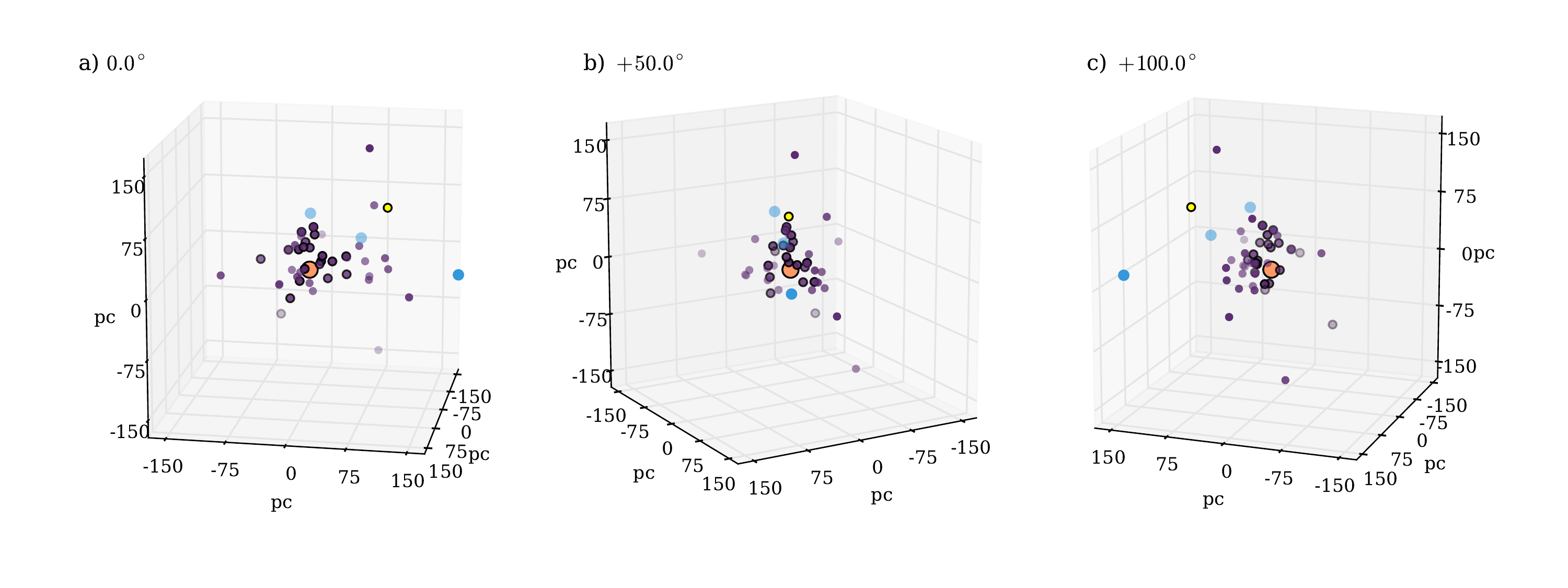}
\caption{The top 50 stars contributing to the external UV field of IM Lup (for $d=161$ pc) as determined using the {\it Hipparcos} data and an interstellar extinction of $A_{\rm UV}~d^{-1}=2$~mag~kpc$^{-1}$. IM Lup is shown in the center as orange and the sun is shown in yellow. Magenta/purple circles correspond to B-stars, blue to O-stars. Stars that have a black outline (besides the sun) are known young stars, while the rest are field stars. \label{fig:externalG0}}
\end{centering}
\end{figure*}
\begin{deluxetable}{cccc}{bh!}
\tablecolumns{4}
\tablewidth{0pt}
\tablecaption{$G_0$ calculated for different line-of-sight $A_{\rm UV}$ values. \label{tab:g0}}
\tabletypesize{\footnotesize}
\tablehead{$A_{\rm UV}$~kpc$^{-1}$ & Field $G_0$ & Young Cluster $G_0$ & Total $G_0$ \\}
\startdata
0 & 2.84 & 1.70 & 4.54\\
1 & 2.38 & 1.47   &3.85 \\
2 & 2.02  & 1.28 & 3.30 \\
3 &  1.75 & 1.11 & 2.86 
\enddata
\end{deluxetable}

\end{document}